%% file: paper.tex
\documentclass[sigconf, screen=true, usenames, dvipsnames, table]{acmart}
\setcopyright{none}

\input{preamble}
\usepackage{graphbox}
\usepackage{svg}
\usepackage{nicefrac}
\usepackage{xfrac}
\usepackage{xspace}
\usepackage{makecell}
\usepackage{bbding}
\usepackage{xcolor,colortbl}
\usepackage{diagbox}

\DeclareMathAlphabet{\mathbcal}{OMS}{cmsy}{b}{n}
\usepackage{amsmath}
\usepackage{mathrsfs}
\usepackage{comment}

\usepackage{bm}
\usepackage{bbm}
\usepackage{enumitem}
\usepackage{hyperref}
\usepackage{cleveref}
\Crefname{Definition}{Definition}{Definitions}
\Crefname{Problem}{Problem}{Problems}

\DeclareRobustCommand{\abbrevcrefs}{\crefname{figure}{fig.}{figs.}\crefname{equation}{Eq.}{Eqs.}\crefname{algorithm}{Alg.}{Algs.}\crefname{section}{Sec.}{Secs.}\crefname{appendix}{App.}{App.}}
\DeclareRobustCommand{\cshref}[1]{{\abbrevcrefs\cref{#1}}}

\usepackage{caption}
\usepackage{subcaption}
\usepackage{soul}

\usepackage{algorithm}
\usepackage{algorithmicx}
\usepackage{algpseudocode}
\algblockdefx[parallel]{ParFor}{EndPar}[1][]{$\textbf{parallel for}$ #1 $\textbf{do}$}{$\textbf{end parallel}$}
\algrenewcommand{\alglinenumber}[1]{\fontsize{6.5}{7}\selectfont#1}
\algtext*{EndPar}

\algblockdefx[parallel]{parfor}{endpar}[1][]{$\textbf{parallel for}$ #1 $\textbf{do}$}{$\textbf{end parallel}$}
\algrenewcommand{\alglinenumber}[1]{\scriptsize#1:}

\algblockdefx[parallel]{parallelfor}{parallelend}
[1][]{\textbf{parallel for} #1}
{\textbf{end parallel}}

\algrenewcommand{\algorithmiccomment}[1]{\hfill$\triangleright$\textit{#1}}

\usepackage{nicefrac}

\algnotext{EndFor}
\algnotext{EndIf}
\algnotext{EndProcedure}

\algnewcommand\algorithmicforeach{\textbf{for each}}
\algdef{S}[FOR]{ForEach}[1]{\algorithmicforeach\ #1\ \algorithmicdo}

\newcommand{\algmargin}{\the\ALG@thistlm}
\newlength{\whilewidth}
\algdef{SE}[parWHILE]{parWhile}{EndparWhile}[1]
{\parbox[t]{\dimexpr\linewidth-\algmargin}{\hangindent\whilewidth\strut\algorithmicwhile\ #1\ \algorithmicdo\strut}}{\algorithmicend\ \algorithmicwhile}\algnewcommand{\parState}[1]{\State \parbox[t]{\dimexpr\linewidth-\algmargin}{\strut #1\strut}}

\newcommand{\setAlgFontSize}{\fontsize{9pt}{10pt}\selectfont} 

\newcommand{\multilinenospace}[1]{\State \parbox[t]{\dimexpr\linewidth-\algorithmicindent}{\begin{spacing}{1.1}\setAlgFontSize#1\strut \end{spacing}}}
\newcommand{\multilinenospaceD}[1]{\State \parbox[t]{\dimexpr0.96 \linewidth-\algorithmicindent}{\begin{spacing}{1.1}\setAlgFontSize#1\strut \end{spacing}}}

\AtBeginDocument{\providecommand\BibTeX{{\normalfont B\kern-0.5em{\scshape i\kern-0.25em b}\kern-0.8em\TeX}}}

\copyrightyear{2022}
\acmYear{2022}
\setcopyright{rightsretained}
\acmConference[WWW '22]{Proceedings of the ACM Web Conference 2022}{April 25--29, 2022}{Virtual Event, Lyon, France.}
\acmBooktitle{Proceedings of the ACM Web Conference 2022 (WWW '22), April 25--29, 2022, Virtual Event, Lyon, France}
\acmDOI{10.1145/3485447.3512160}
\acmISBN{978-1-4503-9096-5/22/04}

\begin{document}

\title[\method: Contrastive Graph Clustering for Community Detection and Tracking]{\method: Contrastive Graph Clustering for\texorpdfstring{\\}{ }Community Detection and Tracking}

\author{\fontsize{11.5pt}{12pt}\selectfont Namyong Park$^{1}$, Ryan Rossi$^2$, Eunyee Koh$^2$, Iftikhar Ahamath Burhanuddin$^2$, Sungchul Kim$^2$, Fan Du$^2$, Nesreen Ahmed$^3$, Christos Faloutsos$^{1}$}
\email{{namyongp,christos}@cs.cmu.edu,{ryrossi,eunyee,burhanud,sukim,fdu}@adobe.com,nesreen.k.ahmed@intel.com}
\affiliation{\institution{\textsuperscript{1}Carnegie Mellon University, \textsuperscript{2}Adobe Research, \textsuperscript{3}Intel Labs}
	\country{}}

\renewcommand{\shortauthors}{N. Park et al.}

\pdfinfo{
/Author (Namyong Park, Ryan Rossi, Eunyee Koh, Iftikhar Ahamath Burhanuddin, Sungchul Kim, Fan Du, Nesreen Ahmed, Christos Faloutsos)
}

\begin{abstract}
Given entities and their interactions in the web data, which may have occurred at different time,
how can we find communities of entities and track their evolution?
In this paper, we approach this important task from graph clustering perspective.
Recently, state-of-the-art clustering performance in various domains has been achieved by deep clustering methods.
Especially, deep graph clustering (DGC) methods have successfully extended deep clustering to graph-structured data
by learning node representations and cluster assignments in a joint optimization framework.
Despite some differences in modeling choices (\eg, encoder architectures), 
existing DGC methods are mainly based on autoencoders and 
use the same clustering objective with relatively minor adaptations.
Also, while many real-world graphs are dynamic,
previous DGC methods considered only static graphs.
In this work, we develop \method, a novel end-to-end framework for graph clustering, which fundamentally differs from existing methods.
\method learns node embeddings and cluster assignments in a contrastive graph learning framework, 
where positive and negative samples are carefully selected in a multi-level scheme
such that they reflect hierarchical community structures and network homophily. 
Also, we extend \method for time-evolving data, 
where temporal graph clustering is performed in an incremental learning fashion, with the ability to detect change points.
Extensive evaluation on real-world graphs demonstrates that 
the proposed \method consistently outperforms existing methods.
\vspace{-2.0em}
\end{abstract}

\ccsdesc[500]{Information systems~Clustering}
\ccsdesc[300]{Information systems~Temporal data}
\ccsdesc[500]{Information systems~Web mining}
\ccsdesc[500]{Computing methodologies~Neural networks}

\keywords{community detection and tracking,
deep graph clustering,
temporal graph clustering,
contrastive learning,
deep graph learning
}

\maketitle

\vspace{-0.75em}
{\fontsize{8pt}{8pt} \selectfont
\textbf{ACM Reference Format:}\\
Namyong Park, Ryan Rossi, Eunyee Koh, Iftikhar Ahamath Burhanuddin, Sungchul Kim, Fan Du, Nesreen Ahmed, Christos Faloutsos. 2022.
CGC: Contrastive Graph Clustering for Community Detection and Tracking.
In \textit{Proceedings of the ACM Web Conference 2022 (WWW ’22), April 25–29, 2022, Virtual Event, Lyon, France.} ACM, New York, NY, USA, 12 pages.
\url{https://doi.org/10.1145/3485447.3512160}}

\input{010introduction}

\input{020problem}

\input{030preliminaries}

\input{040framework}

\input{050experiments}

\input{060relatedwork}

\input{070conclusion}

\bibliographystyle{ACM-Reference-Format}

\clearpage
\appendix

\input{080appendix}
\end{document}

%% file: preamble.tex
\usepackage{xcolor}
\definecolor{thedarkblue}{RGB}{0,0,120} \definecolor{mydarkblue}{rgb}{0,0.08,0.45} \definecolor{darkblue}{rgb}{0,0.08,180}
\colorlet{TufteRed}{red!80!black}
\definecolor{theblue}{RGB}{0,0,180}
\colorlet{thered}{TufteRed}
      
\usepackage{microtype}
\usepackage{balance}

\usepackage{booktabs}
\usepackage{tabularx}

\usepackage{amsmath,amssymb,amsthm}

\newcommand{\eat}[1]{\ignorespaces}
\usepackage{comment}

\usepackage{ragged2e}
\usepackage{multirow}
\usepackage{microtype}
\usepackage{balance}
\usepackage{setspace}

\newcolumntype{H}{>{\setbox0=\hbox\bgroup}c<{\egroup}@{}}

\newcolumntype{R}[1]{>{\RaggedLeft\arraybackslash}} \newcolumntype{L}[1]{>{\RaggedRight\arraybackslash}}

\newcommand{\eg}{\emph{e.g.}}
\newcommand{\ie}{\emph{i.e.}}

\newtheorem{Definition}{\hspace{-1.2em}\bfseries{Definition}}
\newtheorem{Problem}{\hspace{-1.2em}\bfseries{Problem}}

\AtBeginEnvironment{pmatrix}{\setlength{\arraycolsep}{2pt}}

\providecommand{\mat}[1]{\boldsymbol{\mathrm{#1}}}\renewcommand{\vec}[1]{\boldsymbol{\mathrm{#1}}}

\DeclareMathOperator{\E}{\mathbb{E}}
\DeclareMathOperator{\hugeE}{\mbox{\huge\raise-0.3ex\hbox{E}}}
\DeclareMathOperator{\p}{\mathbb{P}}
\DeclareMathOperator{\hugep}{\mbox{\huge\raise-0.3ex\hbox{$\p$}}}

\providecommand{\set}{\mathcal}

\newcommand{\RR}{\mathbb{R}}

\providecommand{\mC}{\ensuremath{\mat{C}}}

\providecommand{\mF}{\ensuremath{\mat{F}}}

\providecommand{\mH}{\ensuremath{\mat{H}}}

\providecommand{\mW}{\ensuremath{\mat{W}}}

\providecommand{\mPhi}{\ensuremath{\mat{\Phi}}}

\providecommand{\vc}{\ensuremath{\vec{c}}}

\providecommand{\vf}{\ensuremath{\vec{f}}}

\providecommand{\vh}{\ensuremath{\vec{h}}}

\providecommand{\vphi}{\ensuremath{\vec{\phi}}}

\providecommand{\numNodes}{n}
\providecommand{\featDim}{d}
\providecommand{\embDim}{d'}

\newcommand{\method}{\textsc{CGC}\xspace}

\newcommand{\acm}{{ACM}\xspace}
\newcommand{\citeseer}{{Citeseer}\xspace}
\newcommand{\dblpS}{{DBLP-S}\xspace}
\newcommand{\magCS}{{MAG-CS}\xspace}

\newcommand{\yahooMsg}{{Yahoo-Msg}\xspace}
\newcommand{\foursquareNYC}{{Foursquare-NYC}\xspace}
\newcommand{\foursquareTKY}{{Foursquare-TKY}\xspace}

\newcommand{\dblpT}{{DBLP-T}\xspace}

\definecolor{Gray}{gray}{0.9}
\definecolor{LightGray}{gray}{0.94}

\newcommand{\midsepremove}{\aboverulesep = 0.5mm \belowrulesep = 0.5mm}

\newcommand{\contribFramework}{\textbf{Novel Framework.}\xspace}
\newcommand{\contribTemporal}{\textbf{Temporal Graph Clustering.}\xspace}
\newcommand{\contribEffective}{\textbf{Effectiveness.}\xspace}

%% file: 010introduction.tex
\begin{figure*}[!t]
\par\vspace{-1.3em}\par
\centering
\makebox[1.0\textwidth][c]{
	\includegraphics[width=0.99\linewidth]{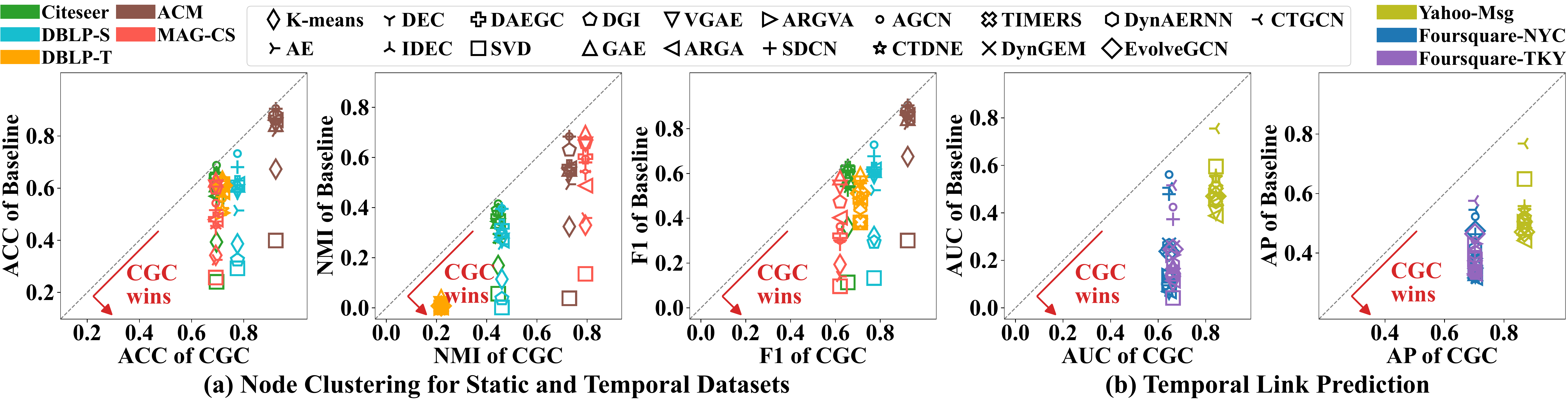}
}
\setlength{\abovecaptionskip}{-1em}
\caption{\underline{\method outperforms competition}: 
All points are below the diagonals for all baselines and graphs.
\method achieves~more accurate (a) node clustering on static and temporal data,
and (b) link prediction based on the time-evolving cluster~{membership}.
}
\label{fig:crownjewel}
\vspace{-1.3em}
\end{figure*}

\vspace{-0.9em}
\section{Introduction}\label{sec:intro}
\vspace{-0.5em}

Given events between two entities, how can we effectively find communities of entities in an unsupervised manner? 
Also, when the events are associated with time, how can we detect communities and track their evolution?
Various web platforms, including social networks, generate data that represent events between entities, 
occurring at a certain time, \eg, check-in records and user interaction logs.
Finding communities from such dyadic temporal events can be formulated as a graph clustering problem,
in which the goal is to find node clusters from a graph,
where the two entities of an event are nodes, and the event forms a temporal edge between them.

\begin{table}[!t]\centering
\par\vspace{-0.2em}\par
\setlength{\abovecaptionskip}{0.1em}
\captionsetup{width=1.0\linewidth}
\caption{
\underline{\method wins on features.} 
	Comparison of the proposed \method with deep learning approaches for graph clustering.
[A]: Aware~of/Utilizing. CL: Clustering, RP: Representation.
}
\small
\makebox[0.4\textwidth][c]{
	\resizebox{0.50\textwidth}{!}{\setlength{\tabcolsep}{0.3mm}
		\renewcommand{\arraystretch}{1} \midsepremove
		\begin{tabular}{lcccccc >{\columncolor{LightGray}}c}\toprule
\diagbox{\textit{Desiderata}}{\textit{Methods}}
			&\makecell[c]{AE\\\cite{hinton2006reducing}} &\makecell[c]{GAE\\\cite{DBLP:journals/corr/KipfW16a}} &\makecell[c]{DAERNN\\\cite{DBLP:journals/kbs/GoyalCC20}} &\makecell[c]{DAEGC\\\cite{DBLP:conf/ijcai/WangPHLJZ19}} &\makecell[c]{SDCN\\\cite{DBLP:conf/www/Bo0SZL020}} &\makecell[c]{AGCN\\\cite{DBLP:conf/mm/PengLJH21}} &\makecell[c]{\textbf{\method}\\\textbf{(Ours)}} \\\midrule
			Jointly optimizing CL and RP & & & &\checkmark &\checkmark &\checkmark &\Checkmark \\
			{[}A{]} Input node features &\checkmark &\checkmark & &\checkmark &\checkmark &\checkmark &\Checkmark \\
			{[}A{]} Network homophily & &\checkmark &\checkmark &\checkmark &\checkmark &\checkmark &\Checkmark \\
			{[}A{]} Hierarchical communities & & & & & & &\Checkmark \\
			Temporal graph clustering & & &\checkmark & & & &\Checkmark\\ \bottomrule\toprule
			\textit{Learning Objective} & & & & & & & \\\midrule
			Contrastive learning-based & & & & & & & \scalebox{0.7}{$\blacksquare$} \\
			Reconstruction-based &\scalebox{0.7}{$\blacksquare$} &\scalebox{0.7}{$\blacksquare$} &\scalebox{0.7}{$\blacksquare$} &\scalebox{0.7}{$\blacksquare$} &\scalebox{0.7}{$\blacksquare$} &\scalebox{0.7}{$\blacksquare$} & \\
			\bottomrule
		\end{tabular}
	}
}
\label{tab:salesmatrix}
\vspace{-4.0em}
\end{table}

In recent years, state-of-the-art clustering performance has been achieved by 
deep clustering methods in several application domains~\cite{DBLP:conf/icml/XieGF16,DBLP:conf/ijcai/GuoGLY17,DBLP:conf/icml/YangFSH17,DBLP:conf/ijcnn/YaoZZHB17,DBLP:conf/bigdataconf/YueLYACS19,DBLP:journals/corr/abs-1802-01059,DBLP:journals/taslp/LuDZ19}.
Following this success, deep graph clustering (DGC)~\cite{DBLP:conf/ijcai/WangPHLJZ19,DBLP:conf/www/Bo0SZL020,DBLP:journals/tcyb/PanHFLJZ20,DBLP:conf/mm/PengLJH21,DBLP:conf/aaai/TianGCCL14} has been receiving increasing attention recently,
which aims to learn cluster-friendly representations using deep neural networks for graph clustering.
Early DGC methods~\cite{DBLP:conf/aaai/TianGCCL14,DBLP:journals/corr/KipfW16a} have taken a two-stage approach, 
where representation learning and clustering are done in isolation;
\eg, node embeddings are learned by graph autoencoders (GAEs)~\cite{DBLP:journals/corr/KipfW16a}, to which a clustering method is applied.
More accurate clustering results have been obtained by another group of DGC methods~\cite{DBLP:conf/ijcai/WangPHLJZ19,DBLP:conf/www/Bo0SZL020,DBLP:conf/mm/PengLJH21}
that adopt a joint optimization framework,
where a clustering objective is combined with the representation learning objective, and both are optimized simultaneously in an end-to-end~manner.

In DGC methods, a major challenge lies in how to effectively utilize node features and graph structure.
Graph neural networks provide an effective framework to this end, 
which propagate and aggregate node features over the graph,
thus learning node embeddings that reflect network homophily.
Further, to make the most of graph structure and node features,
existing methods tried different modeling choices, 
\eg, in terms of encoder architectures (GAEs, attentional GAEs, GAEs with autoencoders (AEs)) and 
how graph structural features and node attributes are combined.
Still, differences among them are relative small: 
They mainly (1) \mbox{perform} reconstruction loss minimization for unsupervised representation learning
(reconstructing the adjacency matrix, node attribute matrix, or both) in an AE-based framework, and
(2) employ the clustering objective first proposed in DEC~\cite{DBLP:conf/icml/XieGF16},
which optimizes cluster assignments by learning from the model's high confidence \mbox{predictions}.

In addition, while many real-world networks are dynamic in nature,
no DGC methods are designed for clustering time-evolving graphs to our knowledge.
Although we can apply existing methods to cluster temporal graphs
(\eg, by ignoring time and applying them to the cumulative graph anew at each time step), 
practical solutions for temporal graph clustering should be able to 
incrementally learn changing community structures, and detect major change points, 
which cannot be addressed effectively by existing~methods.

In this paper, we develop \method, a new graph clustering framework based on contrastive learning, 
which significantly differs from existing DGC methods as summarized in Table~\ref{tab:salesmatrix}.
The main idea of contrastive learning~\cite{DBLP:journals/corr/abs-1807-03748,DBLP:conf/icml/ChenK0H20,DBLP:journals/corr/abs-2004-11362} is 
to pull an entity (called an anchor) and its positive sample closer to each other in the embedding space, 
while pushing the anchor away from its negative sample.
When no labels are available, the choice of positive and negative samples plays a crucial role in contrastive learning.
In such cases, positive samples are often obtained by taking different views of the data 
(\eg, via data augmentations such as rotation and color distortion for images~\cite{DBLP:conf/icml/ChenK0H20}), 
while negative samples are randomly selected from the entire pool of samples.
In \method, based on our understanding of real-world networks and their characteristics 
(\eg, homophily and hierarchical community structures),
we design a multi-level scheme to choose positive and negative samples 
such that they reflect the underlying hierarchical communities and their semantics.
Also, from information theoretic perspective, our contrastive learning objective is designed to maximize the mutual information
between an entity and the hierarchical communities it belongs to in the latent space.
Then guided by this multi-level contrastive objective, 
cluster memberships and entity embeddings are iteratively optimized in an end-to-end framework.

Furthermore, to find communities from time-evolving data, we extend \method framework to the temporal graph clustering setting.
Upon the arrival of new events, entity representations and cluster memberships are updated to reflect the new information, and 
at the same time, temporal smoothness assumption is incorporated into the GNN encoder, and also into the contrastive learning objective,
which enables \method to adapt to changing community structures in a controlled manner.
We also show how \method can be applied to detect major changes occurring in the network, and 
thereby adaptively choose homogeneous historical events to find communities from.

In summary, the key contributions of this work are as follows.
\begin{itemize}[leftmargin=1em,topsep=0pt]
\item \contribFramework
We propose \method, a new contrastive graph clustering framework.
As discussed above and summarized in~\Cref{tab:salesmatrix},
\method is a significant departure from previous DGC methods.

\item \contribTemporal
We extend our \method framework for temporal data.
\method is the first deep graph clustering method for clustering time-evolving networks.

\item \contribEffective
We show the effectiveness of \method via extensive experiments on several static and temporal datasets (\Cref{fig:crownjewel}).

{\setlength{\parindent}{-1.1em}
We release datasets at \url{https://github.com/NamyongPark/CGC-Data}.
\vspace{-1.0em}}

\end{itemize}

%% file: 020problem.tex
\vspace{-1.0em}
\section{Problem Formulation}
\vspace{-0.3em}

In this section, we introduce notations and definitions, and present the problem formulation.
\Cref{tab:symbols} lists the symbols used in~this~work.
\vspace{-2.0em}

\subsection{Graph Clustering}
Let $ G=(V, E) $ be a graph with nodes $ V = \{1,\ldots,n\} $ and 
edges $ E = \{ (u_i, v_i) \mid u_i, v_i \in V \}_{i=1}^{m} $.
Let $ \mF \in \RR^{n \times d} $ be an input node feature matrix.
Let $ k $ denote the number of node clusters.
We define cluster membership as follows to represent node-to-cluster assignment.

\begin{Definition}[Cluster Membership]\label{def:membership}\normalfont
A cluster membership $ \vphi_u\!\in\!\RR_{\ge 0}^{k} $ of node $ u $ is a stochastic vector that adds up to one,
where the $ i $-th entry is the probability of node $ u $ belonging to $ i $-th cluster.
\end{Definition}
\noindent
According to~\Cref{def:membership}, a node belongs to at least one cluster, and can belong to multiple clusters.
Note that this soft cluster membership includes hard cluster assignments as a special case, in which one node belongs to exactly one cluster.
Based on this definition, graph clustering problem is formally defined as follows.
\begin{Problem}[Graph Clustering]\label{prob:graph_clustering}\normalfont
Given a graph $ G=(V, E) $ and input node features $ \mF \in \RR^{n \times d}$, 
learn a cluster membership matrix $ \mPhi \in \RR_{\ge 0}^{n \times k} $ for all $ n $ nodes in $ G $.
\end{Problem}

\noindent
After graph clustering, we want the nodes to be grouped 
such that nodes are more similar to those in the same cluster 
(\eg, in terms of external node labels if available, or connectivity patterns, node features, and structural roles)
than nodes in different clusters.

\vspace{-0.8em}
\subsection{Temporal Graph Clustering}
\vspace{-0.2em}

Let $ G_{\tau}=(V, E_{\tau}) $ be a temporal graph snapshot with nodes $ V = \{1,\ldots,n\} $ and 
temporal edges $ E_{\tau} = \{ (u, v, t) \mid u, v \in V, ~ t \in \tau \} $, 
where $ t $ is time (\eg, a timestamp at the level of milliseconds), and 
$ \tau $ denotes some time span (\eg, one minute, one hour).

\begin{Definition}[Temporal Graph Stream]\label{def:temporal_graph_stream}\normalfont
A temporal graph stream $ \mathcal{G} $ is a sequence of graph snapshots
$ \mathcal{G} = \{ G_{\tau_i} \}_{i=1}^{T} $ where $ T $ is the number of graph snapshots thus far in the stream.
Graph snapshots $ \{ G_{\tau_i} \} $ are assumed to be non-overlapping and ordered in increasing order of time.
\end{Definition}

\begin{Problem}[Temporal Graph Clustering]\label{prob:temporal_graph_clustering}\normalfont
Given a temporal graph stream $ \mathcal{G} = \{ G_{\tau_i} \}_{i=1}^{T} $ and input node features $ \mF \in \RR^{n \times d} $, 
learn a cluster membership matrix $ \mPhi_{i} \in \RR_{\ge 0}^{n \times k} $ for each time span~$ \tau_i $.
\end{Problem}

%% file: 030preliminaries.tex
\section{Preliminaries}

\textbf{Mutual Information (MI) and Contrastive Learning.}
The MI between two random variables (RVs) measures the amount of information obtained about one RV by observing the other RV.
Formally, the MI between two RVs $ X $ and $ Y $, denoted $ I(X;Y) $, is defined as
\begin{align}
	I(X;Y) = \E_{p(x,y)} \left[ \log (\nicefrac{p(x,y)}{p(x)p(y)}) \right]
\end{align}
where $ p(x, y) $ is the joint density of $ X $ and $ Y $, and 
$ p(x) $ and $ p(y) $ denote the marginal densities of $ X $ and $ Y $, respectively.
Several recent studies~\cite{DBLP:conf/iclr/VelickovicFHLBH19,DBLP:journals/corr/abs-1807-03748,DBLP:conf/icml/ChenK0H20,DBLP:conf/icml/BelghaziBROBHC18,DBLP:conf/iclr/HjelmFLGBTB19} 
have seen successful results in representation learning 
by maximizing the MI between a learned representation and different aspects of the data.

Since it is difficult to directly estimate MI~\cite{DBLP:conf/icml/PooleOOAT19}, 
MI maximization is normally done by deriving a lower bound on MI and maximizing it instead.
Intuitively, several lower bounds on MI are based on the idea that 
RVs $ X $ and $ Y $ have a high MI if samples drawn from their joint density $ p(x,y) $ and 
those drawn from the product of marginals $ p(x) p(y) $ can be distinguished accurately.
InfoNCE~\cite{DBLP:journals/corr/abs-1807-03748} is one such lower bound of MI 
in the form of a noise contrastive estimator~\cite{DBLP:journals/jmlr/GutmannH10}:
\begin{align}
I(X;Y) \ge \E \left[ \frac{1}{K} \sum_{i=1}^{K} \log \frac{\exp(f(x_i,y_i))}{\frac{1}{K} \sum_{j=1}^{K} \exp(f(x_i, y_j)) } \right] \triangleq I_{\text{NCE}}(X;Y)
\end{align}
where the expectation is over $ K $ independent samples $ \{ x_i, y_i \}_{i=1}^K $ from the joint density $ p(x,y) $.
Given a set of $ K $ independent samples, the critic function $ f(\cdot) $ aims to predict for each $ x_i $
which one of the $ K $ samples $ x_i $ was drawn together with, \ie, by assigning a large score to the positive pair $ (x_i, y_i) $, and 
small scores to other negative pairs~$ \{ (x_i, y_j) \}_{j \ne i}^K $.

\textbf{Graph Neural Networks (GNNs).}
GNNs are a class of deep learning architectures for graphs that produce node embeddings by repeatedly aggregating local node neighborhoods.
In general, a GNN encoder $ \mathcal{E} $ maps a graph $ G $ and input node features $ \mF \in \RR^{n \times d} $ into 
node embeddings $ \mH \in \RR^{n \times d'} $, that is, $ \mathcal{E}(G, \mF) = \mH $.

%% file: 040framework.tex
\section{Proposed Framework} \label{sec:approach}

In this section, we present the \method framework.
We describe how \method performs graph clustering in a multi-level contrastive learning framework (\Cref{sec:framework:cgc}), and 
discuss how we extend \method for temporal graph clustering (\Cref{sec:framework:temporal}).

\subsection{\method: Contrastive Graph Clustering}\label{sec:framework:cgc}

The proposed framework \method performs contrastive graph clustering by carrying out the following two steps in an alternating fashion:
(1) refining cluster memberships based on the current node embeddings, and 
(2) optimizing node embeddings such that nodes from the same cluster are closer to each other, 
while those from different clusters are pushed further away from each other.

\subsubsection{Multi-Level Contrastive Learning Objective.}\label{sec:framework:cgc:objective}

In \method, contrastive learning happens in the second step above, where 
positive samples of a node are assumed to have been generated by the same cluster as the node of interest,
whereas negative samples are assumed to belong to different clusters.
While no cluster membership labels are available, 
there exist several signals at different levels of the input data 
that we can utilize to effectively construct positive and negative samples for contrastive graph clustering, 
namely, input node features and the characteristics of real-world networks, such as network homophily and hierarchical community structure.

\begin{table}[!t]
	\par\vspace{-0.5em}\par
	\small
	\setlength{\abovecaptionskip}{0.5em}
	\caption{Table of symbols.}
	\centering
	\makebox[0.4\textwidth][c]{
\begin{tabular}{c l}
			\toprule
			\textbf{Symbol} & \multicolumn{1}{c}{\textbf{Definition}} \\
			\midrule
			$ u, v $ & node indices \\
			$ n $ & number of nodes \\
			$ k $ & number of clusters \\
			$ t $ & timestamp of an edge, $ t \ge 0 $ \\
			$ \tau $ & time span \\
			$ G = (V, E) $ & static graph with nodes $ V $ and edges $ E $ \\
			$ \vphi_u \in \RR_{\ge 0}^{k} $ & cluster membership vector of node $ u $ for graph $ G $ \\
$ G_{\tau} = (V, E_{\tau}) $ & \makecell[l]{temporal graph snapshot with nodes $ V $ and\\ temporal edges for time span $ \tau $} \\
			$ \mathcal{G} = \{ G_{\tau_i} \} $ & temporal graph stream  \\
			$ \mPhi_i \in \RR_{\ge 0}^{n \times k} $ & cluster membership matrix for time span $ \tau_i $ \\
			$ \mF \in \RR^{n \times d} $ & input node feature matrix \\
			$ \mH \in \RR^{n \times d'} $ & node embedding matrix \\
			$ \mathcal{N}(u) ~ (\mathcal{N}_\Delta(u)) $ & neighbors of node $ u $ (participating in triangles with $ u $) \\
			$ \set{K}\!=\!\{ k_{\ell} \}_{\ell=1}^{L} $ & number of clusters for contrastive learning \\
			
\bottomrule
		\end{tabular}
	}
	\label{tab:symbols}
	\vspace{-1.5em}
\end{table}

\textbf{Signal: Input Node Features.}
Entities in the same community tend to have similar attributes.
Thus informative node features can be used to distinguish nodes in the same class from those in different classes.
Node features are especially helpful for sparse graphs, since they can complement the scarce relational information.

Therefore, for node $ u $, we take its input features $ \vf_u $ as its positive sample, 
and randomly select another node $ v $ to take its input features~$ \vf_{v} $ as a negative sample;
these positive and negative samples are then contrasted with node embedding $ \vh_u $.
Let $ \set{S}_u^F = \{ \vf_{u}^{\prime i} \}_{i=0}^r $ be the set of one positive ($ i = 0 $) and 
$ r $ negative ($ 1 \le i \le r $) samples (\ie, input features) for node $ u $, where $ \prime $ indicates that sampling was involved.
Since input features and latent embeddings can have different dimensionality, 
we define a node feature-based contrastive loss $ \mathcal{L}_F $
using a bilinear critic parameterized by $ \mW_F \in \RR^{d' \times d} $:
\begin{align}\label{eq:loss:features}
\mathcal{L}_{F}
= \sum_{u=1}^{n} - \log \frac{\exp( (\vh_u^{\intercal} \, \mW_F \, \vf_{u}^{\prime 0}) / \tau)}{\sum_{v=0}^{r} \exp( (\vh_u^{\intercal} \, \mW_F \, \vf_{u}^{\prime v}) / \tau)} 
\end{align}
where $ \tau > 0 $ is a temperature hyper-parameter.

\textbf{Signal: Network Homophily.} In real-world graphs, similar nodes are more likely to attach to each other than dissimilar ones, and
accordingly, a node is more likely to belong to the same cluster as its neighbors than randomly chosen nodes.
In particular, many real-world networks demonstrate the phenomenon of higher-order label homogeneity,
\ie, the tendency of nodes participating in higher-order structures (\eg, triangles) to share the same label,
which is a stronger signal than being connected by an edge alone.
Thus, we use edges and triangles in constructing positive samples.
Further, \method encodes nodes using GNNs, whose neighborhood aggregation scheme also
enforces an inductive bias for network homophily that neighboring nodes have similar representations.

Let $ \mathcal{N}(u) $ denote the neighbors of node $ u $.
Let $ \mathcal{N}_\Delta(u) $ be node~$ u $'s neighbors that participate in the same triangle as node $ u $; 
thus, $ \mathcal{N}_\Delta(u) \subseteq \mathcal{N}(u)$.
A positive sample for node $ u $ is then chosen from among $ \mathcal{N}(u) $,
with a probability of $ \delta / |\mathcal{N}_\Delta(u)| $ for the neighbor in $ \mathcal{N}_\Delta(u) $, and
a probability of $ (1-\delta) / |\mathcal{N}(u) \setminus \mathcal{N}_\Delta(u)| $ for its other neighbors,
where $ \delta \ge 0 $ determines the weight for nodes in $ \mathcal{N}_\Delta(u) $.
Then the positive sample's embeddings are taken from $ \mH\!=\!\mathcal{E}(G, \mF) $.

To construct negative samples, we design a network corruption function $ \mathcal{C}(G, \mF) $,
which constructs a negative network from the original graph $ G $ and input node features $ \mF $.
Specifically, we define $ \mathcal{C}(\cdot) $ to return corrupted node features $ \widetilde{\mF} $, via row-wise shuffling of~$ \mF $,
while preserving the graph $ G $, \ie, $ \mathcal{C}(G, \mF) = (G, \widetilde{\mF}) $,
which can be considered as randomly relocating nodes over the graph while maintaining the graph structure.
Then negative node embeddings $ \widetilde{\mH} \in \RR^{n \times d'} $ are obtained by applying the GNN encoder to $ G $ and~$ \widetilde{\mF} $, and
$ r $ negative samples and their embeddings are randomly chosen.

Let $ \set{S}_u^H = \{ \vh_{u}^{\prime i} \}_{i=0}^r $ be the set containing the embeddings of one positive ($ i = 0 $) and $ r $ negative ($ 1 \le i \le r $) samples for node $ u $.
In \method, a homophily-based contrastive loss $ \mathcal{L}_{H} $ is defined as:
{\setlength{\abovedisplayskip}{1.0em}
\setlength{\belowdisplayskip}{1.0em}
\begin{align}\label{eq:loss:homophily}
\mathcal{L}_{H}
= \sum_{u=1}^{n} - \log \frac{\exp(\vh_u \cdot \vh_{u}^{\prime 0} / \tau)}{\sum_{v=0}^{r} \exp(\vh_u \cdot \vh_{u}^{\prime v} / \tau)} 
\end{align}}where we use an inner product critic function with a temperature hyper-parameter $ \tau > 0 $, and $ \prime $ denoting that sampling was involved.

\textbf{Signal: Hierarchical Community Structure.}
The above loss terms contrast an entity with other individual entities and their input features,
thereby learning community structure at a relatively low level.
Here, we consider communities at a higher level than before
by directly contrasting entities with communities.

\method represents communities as a cluster centroid vector $ \vc \in \RR^{d'} $ in the same latent space as entities,
so that the distance between an entity and cluster centroids reflects the entity's degree of participation in different communities.
To effectively optimize an entity embedding by contrasting it with communities, cluster centroids need to have been embedded 
such that they reflect the underlying community structures and the semantics of input node features.
While the model's initial embeddings of entities and clusters may not capture such community and semantic structures well,
the above two objectives and the use of GNN encoders in \method
effectively guide the optimization process towards identifying meaningful cluster centroids,
especially in the early stage of model training.

Importantly, real-world networks have been shown to exhibit hierarchical community structures.
To model this phenomenon, we design \method to group nodes into a varying number of clusters.
For example, when we aim to group nodes into three clusters, we may also group the same set of nodes into ten and thirty clusters;
then all clustering results taken together reveal hierarchical community structures in different levels of granularities.

Let $ \set{K}\!=\!\{ k_{\ell} \}_{\ell=1}^{L} $ be the set of the number of clusters, and 
$ \mC_{\ell} \in \RR^{k_{\ell} \times d'} $ be the cluster centroid matrix for each $ \ell $.
Given the current node embeddings $ \mH $ and cluster centroids $ \{ \mC_{\ell} \}_{\ell=1}^{L} $,
positive samples for node $ u $ are chosen to be the $ L $ cluster centroids that node~$ u $ most strongly belongs to, 
while its negative samples are randomly selected from among the other $ k_{\ell} - 1 $ cluster centroids for each $ \ell $.
Let $ \set{S}_{u,\ell}^C = \{ \vc_{u,\ell}^{\prime i} \}_{i=0}^{r_\ell} $ be the set 
with the embeddings of one positive ($ i = 0 $) and $ r_\ell $ negative ($ 1 \le i \le r_\ell $) samples (\ie, centroids)
for node $ u $ chosen among $ k_\ell $ centroids.
Using an inner product critic, \method defines a hierarchical community-based contrastive loss $ \mathcal{L}_{C} $ to be:
\begin{align}\label{eq:loss:community}
\mathcal{L}_{C} = \sum_{u=1}^{n} - \left( \frac{1}{L} \sum_{\ell=1}^{L} \log \frac{\exp(\vh_u \cdot \vc_{u,\ell}^{\prime 0} / \tau)}{\sum_{v=0}^{r_\ell} \exp(\vh_u \cdot \vc_{u,\ell}^{\prime v} / \tau)} \right).
\end{align}

\textbf{Multi-Level Contrastive Learning Objective.}
The above loss terms capture signals on the community structure at multiple levels, 
\ie, individual node features ($ \mathcal{L}_{F} $), neighboring nodes ($ \mathcal{L}_{H} $), and 
hierarchically structured communities ($ \mathcal{L}_{C} $).
\method jointly optimizes
\begin{align}\label{eq:loss:combined}
\mathcal{L} = \lambda_F \mathcal{L}_{F} + \lambda_H \mathcal{L}_{H} + \lambda_C \mathcal{L}_{C}
\end{align}
where $ \lambda_F $, $ \lambda_H $, and $ \lambda_C $ are weights for the loss terms.
Via multi-level noise contrastive estimation, \method maximizes the MI between nodes and 
the communities they belong to in the learned latent~space.

\subsubsection{Encoder Architecture.}

As our node encoder $ \mathcal{E} $, we use 
a GNN with
a mean aggregator,
\begin{align}\label{eq:encoder}
\vh_v^l = \text{ReLU} ( \mW_{G} \cdot \text{MEAN}( \{ \vh_v^{l-1} \} \cup \{ \vh_u^{l-1} \mid \forall u \in \mathcal{N}(v) \} ) )
\end{align}
where node $ v $'s embedding $ \vh_v^l $ from the $ l $-th layer of $ \mathcal{E} $ is obtained 
by averaging the embeddings of node $ v $ and its neighbors from the $ (l\!-\!1) $-th layer, 
followed by a linear transformation $ \mW_{G} $ and ReLU non-linearity;
$ \vh_v^0 $ is initialized to be the input node features~$ \vf_v $.

{
\renewcommand{\setAlgFontSize}{\small}\renewcommand{\multilinenospace}[1]{\State \parbox[t]{\dimexpr\linewidth-\algorithmicindent}{\begin{spacing}{1.0}\setAlgFontSize#1\strut \end{spacing}}}
\renewcommand{\multilinenospaceD}[1]{\State \parbox[t]{\dimexpr0.96 \linewidth-\algorithmicindent}{\begin{spacing}{1.0}\setAlgFontSize#1\strut \end{spacing}}}

\algblockdefx[parallel]{parfor}{endpar}[1][]{$\textbf{parallel for}$ #1 $\textbf{do}$}{$\textbf{end parallel}$}
\algrenewcommand{\alglinenumber}[1]{\fontsize{7.5}{8}\selectfont#1\;\;}
\algtext*{EndWhile}
\begin{figure}[t!]
\par\vspace{-1.0em}\par
\vspace{-3mm}
\begin{center}
	\begin{algorithm}[H]
		\caption{\,\textsf{ContrastiveGraphClustering}}
		\label{alg:cgc}
		\begin{algorithmic}[1]
			\Require graph $ G $, input node features $\mF \in \RR^{\numNodes \times \featDim}$,
			clustering algorithm~$ \Pi $,
			number of clusters $ \set{K}=\{ k_\ell \}_{\ell=1}^L $, refinement interval~$ R $
\Ensure cluster membership matrix $\mPhi \in \RR^{\numNodes \times k_1}_{\ge 0} $, 
			node embedding matrix $\mH \in \RR^{\numNodes \times \embDim}$, 
			cluster centroid matrix $ \mC \in \RR^{k_1 \times \embDim} $
			\smallskip
			
			\While{not $ maxEpoch $ and not converged}
			\State $ \mH = \mathcal{E}(G, \mF) $ \Comment{\cshref{eq:encoder}}

			\If{$epoch ~\%~ R = 0$} \Comment{refine clusters and memberships}
				\For{$\ell = 1$ \textbf{to} $ L $}
				\State $ \mC_\ell, \mPhi_\ell = \Pi(\mH, k_\ell) $ \EndFor
			\EndIf

			\State Calculate loss $ \mathcal{L} $ using $ \mH, \mF, \{ \mC_{\ell} \} $ 	\Comment{\cshref{eq:loss:features,eq:loss:homophily,eq:loss:community,eq:loss:combined}}
			\State Backpropagate and optimize model parameters
			\EndWhile

\State $ \mH = \mathcal{E}(G, \mF) $
			\State $ \mC, \mPhi = \Pi(\mH, k_1) $
			\State \Return $ \mPhi, \mH, \mC $
			
			\smallskip
		\end{algorithmic}
\vspace{-1.mm}
	\end{algorithm}
\end{center}
\vspace{-3em}
\end{figure}
}

\subsubsection{Algorithm.}
\Cref{alg:cgc} shows how (1) cluster memberships and (2) node embeddings are alternately optimized in \method.
(1) Given the current node embeddings $ \mH $ produced by $ \mathcal{E} $ (line 2),
a clustering algorithm $ \Pi $ (\eg, $ k $-means) refines cluster centroids $ \{ \mC_{\ell} \} $ and memberships $ \{ \mPhi_{\ell} \} $ (lines 3-5).
(2) Based on the updated cluster centroids and memberships, \method computes the loss and optimizes model parameters (lines 6-7).
In $ \{ k_{\ell} \} $, we assume that $ k_1 $ is the number of clusters that we ultimately want to identify in the network.

\vspace{-1em}
\subsection{\method for Temporal Graph Clustering}\label{sec:framework:temporal}

As a new graph snapshot $ G_{\tau_{i}} $ arrives in a temporal graph stream $ \set{G} = \{ G_{\tau_1}, \ldots, G_{\tau_{i-1}} \} $, 
node embeddings $ \mH_{i-1} $ and cluster memberships $ \mPhi_{i-1} $ 
that \method learned from the snapshots until $ (i\!-\!1) $-th time span
are incrementally updated to reflect the new information in~$ G_{\tau_{i}} $.
Specifically, given a sequence of graph snapshots, \method merges them into a temporal graph and 
performs contrastive graph clustering, taking the temporal information into account.
We use the notation $ G_{i:j} $ to denote a temporal graph that merges the snapshots $ \{ G_{\tau_i}, \ldots, G_{\tau_{j}} \} $,
\ie, $ G_{i:j} = (V, E_{i:j}) $ where $ E_{i:j} = \bigcup_{o=i}^j E_{\tau_{o}} $.
Below we describe how we extend \method for temporal graph clustering.

\subsubsection{Temporal Contrastive Learning Objective.}
As entities interact with each other, their characteristics may change over time, and such temporal changes normally occur smoothly.
Thus, edges of a node observed across a range of time spans provide 
similar and related temporal views of the node in terms of its connectivity pattern. 
Accordingly, given node~$ u $ for time span $ j $, we take its embedding $ \vh_{u,j-1} $ 
obtained in the previous, $ (j\!-\!1) $-th time span as its positive sample.
To obtain negative samples, we use the same network corruption function used in~\Cref{sec:framework:cgc:objective},
obtaining corrupted node features $ \widetilde{\mF} $, and 
take node $ u $'s embedding from the corrupted node embeddings $ \mathcal{E}({G}_{i:j-1}, \widetilde{\mF}) $ as the negative sample;
multiple negative samples can be obtained by using multiple sets of corrupted node features.
Let $ \set{S}_{u,j}^T = \{ \vh_{u,j-1}^{\prime i} \}_{i=0}^r $ be the set 
with the embeddings of one positive ($ i = 0 $) and $ r $ negative ($ 1 \le i \le r $) samples of node $ u $ for the $ j $-th time span,
again $ \prime $ denoting the involvement of sampling.
\method defines a time-based contrastive loss $ \mathcal{L}_{T} $ for time span $ j $ to be:
\begin{align}\label{eq:loss:temporal}
	\mathcal{L}_{T}
	= \sum_{u=1}^{n} - \log \frac{\exp(\vh_{u,j} \cdot \vh_{u,j-1}^{\prime 0} / \tau)}{\sum_{v=0}^{r} \exp(\vh_{u,j} \cdot \vh_{u,j-1}^{\prime v} / \tau)} 
\end{align}
Note that \Cref{eq:loss:temporal} is combined with the objectives discussed in~\Cref{sec:framework:cgc:objective} with a weight of $ \lambda_T $,
augmenting the loss $ \mathcal{L} $ to be
\begin{align}\label{eq:loss:combined:temporal}
	\mathcal{L} = \lambda_F \mathcal{L}_{F} + \lambda_H \mathcal{L}_{H} + \lambda_C \mathcal{L}_{C} + \lambda_T \mathcal{L}_{T}.
\end{align}

\subsubsection{Encoder Architecture.} 
We extend the GNN encoder such that when it aggregates the neighborhood of a node, 
more weight is given to the neighbors that interacted with the node more recently.
To this end, we adjust the weight of a neighbor based on the elapsed time since its latest interaction.
Let $ t_{(u,v)} $ denote the timestamp of an edge between nodes $ u $ and $ v $, and let
$ t_v^{\text{max}} = \max_{u \in \mathcal{N}(v)} \{ t_{(u,v)} \} $, 
\ie, the most recent timestamp when node $ v $ interacted with its neighbors.
With $ \psi $ denoting a time decay factor between 0 and 1,
we apply time decay to the embedding $ \vh_u $ of neighbor $ u $ as follows:
\begin{align}\label{eq:timedecay}
\mathsf{td}(\vh_u) = \psi^{ t^{\text{max}}_v - t_{(u,v)} } \, \vh_u.
\end{align}
Then for time-aware neighborhood aggregation, $ \vh_u $ in~\Cref{eq:encoder} is replaced with its time decayed version $ \mathsf{td}(\vh_u) $.

\subsubsection{Graph Stream Segmentation.}\label{sec:framework:temporal:segmentation}

Given a new graph snapshot, \method merges it with the previous ones, and 
refines cluster memberships on the resulting temporal graph.
This process is based on the assumption that new events are similar to earlier ones.
However, the new snapshot may differ greatly from the previous ones,
when significant changes have occurred in the network.
Detecting such changes is important, as it lets \method find clusters from snapshots with similar patterns, and 
such events also correspond to important milestones or anomalies in the network.

Let $ \mathcal{G}_{\text{seg}} = \{ G_{\tau_i},\ldots, G_{\tau_j} \}$ be 
the current graph stream segment for some $ i $ and $ j $ $(i < j) $.
Given a new snapshot $ G_{\tau_{j+1}} $,
we expand the current segment $ \mathcal{G}_{\text{seg}} $ with $ G_{\tau_{j+1}} $ 
if $ G_{\tau_{j+1}} $ is similar to $ \mathcal{G}_{\text{seg}} $;
if not, we start a new graph stream segment consisting only of $ G_{\tau_{j+1}} $.
This is basically a binary decision problem on whether to segment the graph stream or not.
Our idea to solve this problem is
to compare the embeddings of the nodes appearing in both $ \mathcal{G}_{\text{seg}} $ and $ G_{\tau_{j+1}} $.
Note that the GNN encoder in this step was trained with the graphs in the observed segment, 
and no further training has been performed on the new snapshot.
Since embeddings from GNNs reflect the characteristics of nodes that \method learned from the existing segment,
the embeddings of the nodes in the new graph $ G_{\tau_{j+1}} $ will be similar to 
their embeddings in the existing segment $ \mathcal{G}_{\text{seg}} $
if $ G_{\tau_{j+1}} $ is similar to $ \mathcal{G}_{\text{seg}} $. 
By the same token, a major change in the new snapshot 
will lead to a large difference between the embeddings of a node in $ G_{\tau_{j+1}} $ and $ \mathcal{G}_{\text{seg}} $.
Let $ V^* $ be the nodes appearing in both $ \mathcal{G}_{\text{seg}} $ and $ G_{\tau_{j+1}} $.
Let $ \mH^{\text{seg}}_{V^*}, ~ \mH^{j+1}_{V^*} \in \mathbb{R}^{|V^*| \times \embDim} $ 
be the two sets of embeddings of the nodes in $ V^* $, 
computed for $ \mathcal{G}_{\text{seg}} $ and $ G_{\tau_{j+1}} $, respectively, as discussed above.
Using a distance metric $ d(\cdot, \cdot) $ (\eg, cosine distance),
we define the distance $ \textsf{Dist}(\cdot,\cdot) $ between $ \mH^{\text{seg}}_{V^*} $ and $ \mH^{j+1}_{V^*} $ to be
\begin{align}
	\textsf{Dist}(\mH^{\text{seg}}_{V^*}, \mH^{t+1}_{V^*}) = \text{MEAN}\{ d((\mH^{\text{seg}}_{V^*})_i, (\mH^{t+1}_{V^*})_i) \mid i \in V^* \}
\end{align}
and segment the stream if the distance is beyond a threshold (\cshref{alg:segmentation}).

{
\renewcommand{\setAlgFontSize}{\small}\renewcommand{\multilinenospace}[1]{\State \parbox[t]{\dimexpr\linewidth-\algorithmicindent}{\begin{spacing}{1.0}\setAlgFontSize#1\strut \end{spacing}}}
\renewcommand{\multilinenospaceD}[1]{\State \parbox[t]{\dimexpr0.96 \linewidth-\algorithmicindent}{\begin{spacing}{1.0}\setAlgFontSize#1\strut \end{spacing}}}

\algblockdefx[parallel]{parfor}{endpar}[1][]{$\textbf{parallel for}$ #1 $\textbf{do}$}{$\textbf{end parallel}$}
\algrenewcommand{\alglinenumber}[1]{\fontsize{7.5}{8}\selectfont#1\;\;}
\begin{figure}[t!]
	\vspace{-6mm}
\begin{center}
		\begin{algorithm}[H]
			\caption{\,\method Framework}
			\label{alg:framework}
			\begin{algorithmic}[1]
				\Require graph stream $ \mathcal{G} $, input node feature matrix $\mF \in \RR^{\numNodes \times \embDim}$
\Ensure $ \big\{ $ cluster memberships $\mPhi_i \in \RR^{\numNodes \times k}$, 
				node embeddings $\mH_i \in \RR^{\numNodes \times \embDim}$, 
graph stream segment $ \mathcal{G}_i^{\text{seg}} \big\} $ for each time span $ i $
				\smallskip
				
				\State $ \mathcal{G}_0^{\text{seg}} = \{\} $
				\ForEach {$G_{\tau_{i}} \in \mathcal{G} $}
				\State $ \mathcal{G}_{i}^{\text{seg}}\!=\!$ \textsf{GraphStreamSegmentation}($ G_{\tau_{i}}, \mathcal{G}_{i-1}^{\text{seg}}, \mF $) \Comment{\cshref{alg:segmentation}}
\State $ \mPhi_i, \mH_i, \mC_i\!=\!$ \textsf{ContrastiveGraphClustering}(\textsf{Merge}($ \mathcal{G}_i^{\text{seg}}$), $ \mF $) \Comment{\cshref{alg:cgc}}
				\EndFor
				
				\State \Return $ \{ \mPhi_i, \mH_i, \mathcal{G}_i^{\text{seg}} \}_i $
				
				\smallskip
			\end{algorithmic}
\vspace{-1.mm}
		\end{algorithm}
	\end{center}
	\vspace{-9mm}
\end{figure}
}

\vspace{-0.5em}
\subsubsection{Putting Things Together.}
\method tracks changing cluster memberships in an incremental end-to-end framework (\cshref{alg:framework}).
As a new graph snapshot arrives, \method adaptively determines 
a sequence of graph snapshots to find clusters from, using \cshref{alg:segmentation} (line~3), 
and updates clustering results and node embeddings, using \cshref{alg:cgc} (line~4).

%% file: 050experiments.tex
\vspace{-0.7em}
\section{Experiments}\label{sec:exp}
\vspace{-0.3em}

The experiments are designed to answer the following questions:
\begin{itemize}[leftmargin=*,topsep=0pt]
\item \textbf{RQ1 (Node Clustering):} 
Given static and temporal graphs, 
how accurately can the proposed \method cluster nodes? (\Cref{sec:exp:nodeclustering})

\item \textbf{RQ2 (Temporal Link Prediction):} 
How informative is the learned cluster membership in predicting temporal links?
(\Cref{sec:exp:linkpred})

\item \textbf{RQ3 (Ablation Study):} 
How do different variants of the proposed framework affect the clustering performance? (\Cref{sec:exp:ablation})

\end{itemize}
Further results are in Appendix, \eg, mining case studies (\cshref{sec:app:casestudy}).

\vspace{-1.0em}
\subsection{Datasets}\label{sec:exp:data}

\subsubsection{Static Datasets.}
\Cref{tab:datasets:static} presents the statistics of static datasets.
These datasets have labels and input features for all nodes.

\textbf{\acm}~\cite{DBLP:conf/www/Bo0SZL020} is a paper network from the ACM digital library~\cite{acm},
where two papers are linked by an edge if they are written by the same author.
Papers in this dataset are published in KDD, SIGMOD, SIGCOMM, and MobiCom, and 
belong to one of the following three classes: database, wireless communication, and data mining.
Node features are the bag-of-words of the paper keywords.

\textbf{\dblpS}~\cite{DBLP:conf/www/Bo0SZL020} is an author network from the DBLP computer science bibliography~\cite{dblp},
where an edge connects two authors (\ie, nodes) if they have a coauthor relationship.
Authors are divided into the following four areas, according to the conferences of their publications: 
database, data mining, machine learning, and information retrieval.
Node features are the bag-of-words of their keywords.

\textbf{\citeseer}~\cite{DBLP:conf/www/Bo0SZL020} is a citation network from the CiteSeer digital library~\cite{citeseer}, 
where an edge denotes a citation between two documents.
Documents are assigned to one of the six areas: 
agents, AI, database, information retrieval, machine language, and human-computer interaction.
Node features are the bag-of-words of the documents.

\textbf{\magCS}~\cite{shchur2018pitfalls} is a network of authors in CS from the Microsoft Academic Graph.
An edge connects two authors (\ie, nodes) if they co-authored a paper.
Node features are keywords of the author's papers, and labels denote the most active field of study of each~\mbox{author}.

\vspace{-0.7em}
\subsubsection{Temporal Datasets.}
\Cref{tab:datasets:temporal} presents the statistics of temporal datasets.
These datasets do not contain input node features, and dynamic node labels are available only for \dblpT.

\textbf{\dblpT}~\cite{DBLP:conf/aaai/YaoJ21} is an author network from DBLP~\cite{dblp},
where edges denote coauthorship from 2004 to 2018.
Node labels represent the authors' research areas (computer networks and machine learning), and
may change over time as authors switch their research focus.

\textbf{\yahooMsg}~\cite{yahoo} is a communication network among Yahoo! Messenger users,
where two users are linked by an edge if a user sent a message to another user.

\textbf{\foursquareNYC} and \textbf{\foursquareTKY}~\cite{DBLP:journals/tsmc/YangZZY15} are 
user check-in records, collected by Foursquare~\cite{foursquare} between April 2012 and February 2013 from New York City and Tokyo, respectively.
An edge links a user and a venue if a user checked in to the venue.

\vspace{-1.2em}
\subsection{Baselines}\label{sec:exp:baselines}
\vspace{-0.2em}

\hspace*{0.8\parindent}
\textbf{Static Baselines.}
K-means~\cite{hartigan1979algorithm} is a classic clustering method applied to the raw input features.
AE~\cite{hinton2006reducing} produces node embeddings by using autoencoders.
DEC~\cite{DBLP:conf/icml/XieGF16} is a deep clustering method that optimizes node embeddings and performs clustering simultaneously.
IDEC~\cite{DBLP:conf/ijcai/GuoGLY17} extends DEC by adding a reconstruction loss. 

A group of methods also take graph structures into account for node representation learning and graph clustering.
SVD~\cite{golub1971singular} applies singular value decomposition to the adjacency matrix.
GAE~\cite{DBLP:journals/corr/KipfW16a} and VGAE~\cite{DBLP:journals/corr/KipfW16a} employ a graph autoencoder and a variational variant.
ARGA~\cite{DBLP:journals/tcyb/PanHFLJZ20} and ARGVA~\cite{DBLP:journals/tcyb/PanHFLJZ20} are 
an adversarially regularized graph autoencoder and its variational version.
DGI~\cite{DBLP:conf/iclr/VelickovicFHLBH19} learns node embeddings by maximizing their MI with the graph.
DAEGC~\cite{DBLP:conf/ijcai/WangPHLJZ19}, SDCN~\cite{DBLP:conf/www/Bo0SZL020}, and AGCN~\cite{DBLP:conf/mm/PengLJH21} are 
deep graph clustering methods that jointly optimize node embeddings and graph clustering.

\textbf{Temporal Baselines.}
CTDNE~\cite{DBLP:conf/www/NguyenLRAKK18} learns node embeddings based on temporal random walks.
TIMERS~\cite{DBLP:conf/aaai/ZhangCPW018} is an incremental SVD method that employs error-bounded SVD restart on dynamic networks.
DynGEM~\cite{DBLP:journals/corr/abs-1805-11273} leverages AEs to incrementally generate node embeddings at time $ t $ 
by using the graph snapshot at time $ t-1 $.
DynAERNN~\cite{DBLP:journals/kbs/GoyalCC20} uses historical adjacency matrices to reconstruct the current one
by using an encoder-decoder architecture with RNNs.
EvolveGCN~\cite{DBLP:conf/aaai/ParejaDCMSKKSL20} models how the parameters of GCNs~\cite{DBLP:conf/iclr/KipfW17} evolve over time.
CTGCN~\cite{CTGCN} is a k-core based temporal GCN.

For methods that produce only node embeddings (\eg, AE, SVD, GAE, CTDNE), 
we apply $ k $-means to the node embeddings to obtain cluster memberships.
As the temporal link prediction task in \Cref{sec:exp:linkpred} involves dot product scores,
we apply Gaussian mixture models to node embeddings to obtain soft cluster memberships.
\Cref{sec:app:settings} presents experimental settings of baselines and \method.

\begin{table*}[!htp]\centering
\par\vspace{-1.3em}\par
\setlength{\abovecaptionskip}{0.2em}
\captionsetup{width=1.05\linewidth}
\caption{\method achieves the best node clustering results on static graphs. 
	Best results are in bold, and second best results are underlined.}
\label{tab:results:nodeclus:static}
\small
\makebox[0.4\textwidth][c]{
\setlength{\tabcolsep}{0.5mm}
\begin{tabular}{lrrrrrrrrrrrrrrrr}\toprule
	\multirow{2}{*}{Method} &\multicolumn{4}{c}{\dblpS} &\multicolumn{4}{c}{\acm} &\multicolumn{4}{c}{\citeseer} &\multicolumn{4}{c}{\magCS} \\
	\cmidrule(l{2pt}r{2pt}){2-5} \cmidrule(l{2pt}r{2pt}){6-9} \cmidrule(l{2pt}r{2pt}){10-13} \cmidrule(l{2pt}r{2pt}){14-17}
	&\makecell[r]{ACC} &\makecell[r]{NMI} &\makecell[r]{ARI} &\makecell[r]{F1} &\makecell[r]{ACC} &\makecell[r]{NMI} &\makecell[r]{ARI} &\makecell[r]{F1} &\makecell[r]{ACC} &\makecell[r]{NMI} &\makecell[r]{ARI} &\makecell[r]{F1} &\makecell[r]{ACC} &\makecell[r]{NMI} &\makecell[r]{ARI} &\makecell[r]{F1} \\\midrule
	K-means~\cite{hartigan1979algorithm} &38.7$\pm$0.7 &11.5$\pm$0.4 &7.0$\pm$0.4 &31.9$\pm$0.3 &67.3$\pm$0.7 &32.4$\pm$0.5 &30.6$\pm$0.7 &67.6$\pm$0.7 &39.3$\pm$3.2 &16.9$\pm$3.2 &13.4$\pm$3.0 &36.1$\pm$3.5 &34.2$\pm$2.2 &33.0$\pm$1.5 &4.5$\pm$1.3 &19.4$\pm$0.4 \\
	AE~\cite{hinton2006reducing} &51.4$\pm$0.4 &25.4$\pm$0.2 &12.2$\pm$0.4 &52.5$\pm$0.4 &81.8$\pm$0.1 &49.3$\pm$0.2 &54.6$\pm$0.2 &82.0$\pm$0.1 &57.1$\pm$0.1 &27.6$\pm$0.1 &29.3$\pm$0.1 &53.8$\pm$0.1 &32.5$\pm$1.9 &35.9$\pm$2.3 &12.9$\pm$1.5 &14.0$\pm$1.1 \\
	DEC~\cite{DBLP:conf/icml/XieGF16} &58.2$\pm$0.6 &29.5$\pm$0.3 &23.9$\pm$0.4 &59.4$\pm$0.5 &84.3$\pm$0.8 &54.5$\pm$1.5 &60.6$\pm$1.9 &84.5$\pm$0.7 &55.9$\pm$0.2 &28.3$\pm$0.3 &28.1$\pm$0.4 &52.6$\pm$0.2 &44.4$\pm$3.4 &53.5$\pm$2.8 &33.6$\pm$4.0 &28.4$\pm$3.1 \\
	IDEC~\cite{DBLP:conf/ijcai/GuoGLY17} &60.3$\pm$0.6 &31.2$\pm$0.5 &25.4$\pm$0.6 &61.3$\pm$0.6 &85.1$\pm$0.5 &56.6$\pm$1.2 &62.2$\pm$1.5 &85.1$\pm$0.5 &60.5$\pm$1.4 &27.2$\pm$2.4 &25.7$\pm$2.7 &61.6$\pm$1.4 &45.7$\pm$1.8 &55.3$\pm$2.6 &33.5$\pm$3.4 &30.8$\pm$2.3 \\
	SVD~\cite{golub1971singular} &29.3$\pm$0.4 &0.1$\pm$0.0 &0.0$\pm$0.1 &13.3$\pm$2.2 &39.9$\pm$5.8 &3.8$\pm$4.3 &3.1$\pm$4.2 &30.1$\pm$8.2 &24.1$\pm$1.2 &5.7$\pm$1.5 &0.1$\pm$0.3 &11.4$\pm$1.7 &25.7$\pm$4.4 &13.6$\pm$7.3 &1.3$\pm$2.2 &9.7$\pm$4.6 \\
	DGI~\cite{DBLP:conf/iclr/VelickovicFHLBH19} &32.5$\pm$2.4 &3.7$\pm$1.8 &1.7$\pm$0.9 &29.3$\pm$3.3 &88.0$\pm$1.1 &63.0$\pm$1.9 &67.7$\pm$2.5 &88.0$\pm$1.0 &64.1$\pm$1.3 &38.8$\pm$1.2 &38.1$\pm$1.9 &60.4$\pm$0.9 &60.0$\pm$0.6 &65.9$\pm$0.4 &50.3$\pm$0.9 &47.3$\pm$0.4 \\
	GAE~\cite{DBLP:journals/corr/KipfW16a} &61.2$\pm$1.2 &30.8$\pm$0.9 &22.0$\pm$1.4 &61.4$\pm$2.2 &84.5$\pm$1.4 &55.4$\pm$1.9 &59.5$\pm$3.1 &84.7$\pm$1.3 &61.4$\pm$0.8 &34.6$\pm$0.7 &33.6$\pm$1.2 &57.4$\pm$0.8 &\ul{63.2$\pm$2.6} &\ul{69.9$\pm$0.6} &\ul{52.8$\pm$1.5} &\ul{58.1$\pm$4.1} \\
	VGAE~\cite{DBLP:journals/corr/KipfW16a} &58.6$\pm$0.1 &26.9$\pm$0.1 &17.9$\pm$0.1 &58.7$\pm$0.1 &84.1$\pm$0.2 &53.2$\pm$0.5 &57.7$\pm$0.7 &84.2$\pm$0.2 &61.0$\pm$0.4 &32.7$\pm$0.3 &33.1$\pm$0.5 &57.7$\pm$0.5 &60.4$\pm$2.9 &65.3$\pm$1.4 &50.0$\pm$2.1 &53.8$\pm$4.0 \\
	ARGA~\cite{DBLP:journals/tcyb/PanHFLJZ20} &61.6$\pm$1.0 &26.8$\pm$1.0 &22.7$\pm$0.3 &61.8$\pm$0.9 &86.1$\pm$1.2 &55.7$\pm$1.4 &62.9$\pm$2.1 &86.1$\pm$1.2 &56.9$\pm$0.7 &34.5$\pm$0.8 &33.4$\pm$1.5 &54.8$\pm$0.8 &47.9$\pm$6.0 &48.7$\pm$3.0 &23.6$\pm$9.0 &40.3$\pm$5.0 \\
	DAEGC~\cite{DBLP:conf/ijcai/WangPHLJZ19} &62.1$\pm$0.5 &32.5$\pm$0.5 &21.0$\pm$0.5 &61.8$\pm$0.7 &86.9$\pm$2.8 &56.2$\pm$4.2 &59.4$\pm$3.9 &87.1$\pm$2.8 &64.5$\pm$1.4 &36.4$\pm$0.9 &37.8$\pm$1.2 &62.2$\pm$1.3 &48.1$\pm$3.8 &60.3$\pm$0.8 &47.4$\pm$4.2 &32.2$\pm$3.2 \\
SDCN~\cite{DBLP:conf/www/Bo0SZL020} &68.1$\pm$1.8 &39.5$\pm$1.3 &39.2$\pm$2.0 &67.7$\pm$1.5 &90.5$\pm$0.2 &68.3$\pm$0.3 &73.9$\pm$0.4 &90.4$\pm$0.2 &66.0$\pm$0.3 &38.7$\pm$0.3 &40.2$\pm$0.4 &\ul{63.6$\pm$0.2} &51.6$\pm$5.5 &58.0$\pm$1.9 &46.9$\pm$8.1 &30.2$\pm$4.3 \\
	AGCN~\cite{DBLP:conf/mm/PengLJH21} &\ul{73.3$\pm$0.4} &\ul{39.7$\pm$0.4} &\ul{42.5$\pm$0.3} &\ul{72.8$\pm$0.6} &\ul{90.6$\pm$0.2} &\ul{68.4$\pm$0.5} &\ul{74.2$\pm$0.4} &\ul{90.6$\pm$0.2} &\ul{68.8$\pm$0.2} &\ul{41.5$\pm$0.3} &\ul{43.8$\pm$0.3} &62.4$\pm$0.2 &54.2$\pm$5.2 &59.4$\pm$2.1 &49.2$\pm$6.5 &36.3$\pm$4.4 \\\midrule\midrule
	\method (Ours) &\textbf{77.6$\pm$0.5} &\textbf{46.1$\pm$0.6} &\textbf{49.7$\pm$1.1} &\textbf{77.2$\pm$0.4} &\textbf{92.3$\pm$0.3} &\textbf{72.9$\pm$0.7} &\textbf{78.4$\pm$0.6} &\textbf{92.3$\pm$0.3} &\textbf{69.6$\pm$0.6} &\textbf{44.6$\pm$0.6} &\textbf{46.0$\pm$0.6} &\textbf{65.5$\pm$0.7} &\textbf{69.3$\pm$4.0} &\textbf{79.3$\pm$1.2} &\textbf{64.4$\pm$3.7} &\textbf{62.1$\pm$4.5} \\
	\bottomrule
\end{tabular}
}
\vspace{-1.2em}
\end{table*}

\begin{table}[!htp]\centering
\par\vspace{-0.5em}\par
\setlength{\abovecaptionskip}{0.2em}
\caption{\method achieves the highest node clustering accuracy on the temporal \dblpT graph. 
Best results are in bold, and second best results are underlined.}
\label{tab:results:nodeclus:temporal}
\small
\makebox[0.4\textwidth][c]{
\renewcommand{\arraystretch}{0.95}
\setlength{\tabcolsep}{1.2mm}
\begin{tabular}{lrrrr}\toprule
	\multirow{2}{*}{Method} &\multicolumn{4}{c}{\dblpT} \\\cmidrule{2-5}
	&\makecell[c]{ACC} &\makecell[c]{NMI} &\makecell[c]{ARI} &\makecell[c]{F1} \\\midrule
	SVD~\cite{golub1971singular} &61.60$\pm$0.01 &0.16$\pm$0.02 &-0.06$\pm$0.01 &38.13$\pm$0.02 \\
	SVD-latest &61.62$\pm$0.02 &0.16$\pm$0.02 &-0.04$\pm$0.02 &38.17$\pm$0.04 \\
	DGI~\cite{DBLP:conf/iclr/VelickovicFHLBH19} &61.64$\pm$0.02 &0.06$\pm$0.01 &0.08$\pm$0.01 &38.77$\pm$0.07 \\
	DGI-latest &61.66$\pm$0.02 &0.06$\pm$0.02 &0.03$\pm$0.02 &38.44$\pm$0.06 \\
	GAE~\cite{DBLP:journals/corr/KipfW16a} &\ul{63.76$\pm$0.18} &\ul{4.40$\pm$0.16} &\ul{7.28$\pm$0.25} &\ul{59.75$\pm$0.20} \\
	GAE-latest &60.17$\pm$0.04 &0.72$\pm$0.02 &2.47$\pm$0.05 &52.36$\pm$0.11 \\
	VGAE~\cite{DBLP:journals/corr/KipfW16a} &60.06$\pm$0.18 &1.63$\pm$0.06 &3.44$\pm$0.11 &55.66$\pm$0.11 \\
	VGAE-latest &60.67$\pm$0.03 &0.77$\pm$0.02 &2.61$\pm$0.03 &51.90$\pm$0.06 \\
	ARGA~\cite{DBLP:journals/tcyb/PanHFLJZ20} &58.46$\pm$0.25 &0.16$\pm$0.04 &0.86$\pm$0.16 &48.95$\pm$0.27 \\
	ARGA-latest &60.54$\pm$0.13 &0.19$\pm$0.05 &0.81$\pm$0.15 &45.37$\pm$0.30 \\
SDCN~\cite{DBLP:conf/www/Bo0SZL020} &56.70$\pm$0.60 &2.18$\pm$0.72 &2.88$\pm$0.51 &55.66$\pm$0.87 \\
	SDCN-latest &51.51$\pm$0.26 &0.13$\pm$0.03 &0.11$\pm$0.04 &50.79$\pm$0.30 \\
	AGCN~\cite{DBLP:conf/mm/PengLJH21} &56.04$\pm$0.86 &0.88$\pm$0.38 &1.11$\pm$0.40 &50.34$\pm$1.13 \\
	AGCN-latest &54.52$\pm$0.91 &0.09$\pm$0.03 &0.14$\pm$0.12 &48.67$\pm$0.85 \\ \midrule
	CTDNE~\cite{DBLP:conf/www/NguyenLRAKK18} &51.58$\pm$0.07 &1.98$\pm$0.06 &-0.99$\pm$0.03 &48.19$\pm$0.27 \\
	CTDNE-latest &50.57$\pm$0.10 &0.02$\pm$0.01 &0.01$\pm$0.01 &49.85$\pm$0.10 \\
	TIMERS~\cite{DBLP:conf/aaai/ZhangCPW018} &61.70$\pm$0.00 &0.09$\pm$0.01 &0.02$\pm$0.00 &38.21$\pm$0.01 \\
	DynGEM~\cite{DBLP:journals/corr/abs-1805-11273} &60.73$\pm$0.12 &0.27$\pm$0.04 &1.26$\pm$0.12 &46.52$\pm$0.22 \\
	DynAERNN~\cite{DBLP:journals/kbs/GoyalCC20} &62.34$\pm$0.09 &0.69$\pm$0.08 &1.66$\pm$0.13 &44.83$\pm$0.22 \\
	EvolveGCN~\cite{DBLP:conf/aaai/ParejaDCMSKKSL20} &61.02$\pm$0.00 &0.79$\pm$0.00 &2.64$\pm$0.00 &51.16$\pm$0.02 \\
	CTGCN~\cite{CTGCN} &59.07$\pm$0.47 &1.06$\pm$0.12 &2.88$\pm$0.27 &55.14$\pm$0.23 \\ \midrule\midrule
	\method (Ours) &\textbf{71.82$\pm$0.99} &\textbf{21.87$\pm$1.85} &\textbf{27.28$\pm$2.93} &\textbf{71.12$\pm$0.86} \\
	\bottomrule
\end{tabular}
}
\vspace{-1.5em}
\end{table}

\vspace{-0.5em}
\subsection{Node Clustering Quality (RQ1)}\label{sec:exp:nodeclustering}

We evaluate the clustering quality using static and temporal graphs with node labels (\citeseer, \dblpS, \acm, \magCS, and \dblpT).
Given cluster assignments, the best match between clusters and node labels is 
obtained by the Munkres algorithm~\cite{kuhn1955hungarian}, and
clustering performance is measured using four metrics, which range from 0 to 1 (higher values are better):
ACC (Accuracy), NMI (Normalized Mutual Information), ARI (Adjusted Rand Index), and F1 score.

\vspace{-0.5em}
\subsubsection{Static Datasets.}

\Cref{tab:results:nodeclus:static} shows the results on static graphs.
The proposed method \method consistently outperforms existing methods on all datasets in four metrics.
Our novel multi-level contrastive graph learning objectives enable \method to accurately identify node clusters 
by effectively leveraging the characteristics of real-world networks. We summarize our observations on the results below.

(1) Deep clustering methods (DEC, IDEC) outperform AE, 
which performs dimensionality reduction of the input features without clustering objectives.
(2) Comparing AE against GAE and ARGA, we can see that utilizing graph structures improves the clustering quality;
in some cases, the performance of GAE and ARGA is even better than DEC and IDEC, although they do not have clustering objectives.
(3) Deep graph clustering methods (DAEGC, SDCN, AGCN) further improve upon deep clustering methods and 
those that learn from input features or the graph structure without clustering objectives,
which shows the benefit of combining deep clustering with graph structural information.
(4) A comparison with DGI is also noteworthy, as DGI learns node embeddings via MI maximization over a graph.
Despite some similarity, DGI cannot effectively identify community structures,
as it maximizes the MI between nodes and the entire graph, without regard to communities therein.

\vspace{-0.5em}
\subsubsection{Temporal Datasets.}\label{sec:exp:nodeclustering:temporal}

Results on the temporal graph \dblpT are in \Cref{tab:results:nodeclus:temporal},
which reports the average of the clustering performance over multiple temporal snapshots.
Since static baselines have no notion of graph stream segmentation, 
it is up to the user to decide which data to provide as input.
We evaluate static baselines in two widely used settings, representative of the way existing temporal graph clustering methods operate:
The default setting is to use all observed snapshots at each time step, and 
the other setting is to use only the latest graph snapshot (marked with ``-latest'' suffix).

\method outperforms all baselines, achieving up to 13\% and 397\% higher ACC and NMI, respectively, than the best performing baseline.
Notably, nearly all baselines do not perform well, obtaining close to zero NMI and ARI,
which demonstrates the difficult of finding clusters over time-evolving networks.
Especially, no input features are available for \dblpT, which poses an additional challenge to methods that heavily rely on them.
For static baselines, using all snapshots often led to similar or better results in comparison to using the last snapshot.
Results also show that temporal baselines fail to identify changing community structure.
While they are designed to keep track of time-evolving node embeddings, 
their representation learning mechanism does not take clustering objective into account,
which makes them less effective for community detection.
\Cref{fig:performance_over_time:node_clustering} in~\Cref{sec:app:performance_over_time} 
shows how ACC and NMI of \method and four select baselines change over time.
While baselines' performance shows an upward trend, their improvement is not significant.
On the other hand, \method's performance improves remarkably over time, 
successfully identifying changing \mbox{communities}.

\vspace{-0.5em}
\subsection{Temporal Link Prediction Accuracy (RQ2)} \label{sec:exp:linkpred}

The task is to predict the graph $G_{\tau'}=(V,E_{\tau'})$ for the next time span $ \tau' $,
where $E_{\tau'}$ are the temporal positive (\ie, observed) edges.
We uniformly randomly sample the same amount of temporal negative edges $ E_{\tau'}^{-} $ such that
$ E_{\tau'}^{-} = \{(u,v) \,|\, 
    u,v \sim \text{Uniform}(1,\ldots,n) \wedge 
(u,v) \not\in E_{\tau'} \}$.
Given an edge $(u,v) \in E_{\tau'} \cup E_{\tau'}^{-}$ for time span $\tau'$ to predict, 
we estimate the likelihood of such an edge existing as 
$ A_{uv}^{\tau'} = \vphi_u^{\intercal} \vphi_v $,
where $\vphi_u$ and $\vphi_v$ are cluster memberships for nodes $u$ and $v$.
We can use link prediction task for evaluating clustering quality, 
since nodes in the same cluster are more likely to form a link between them than nodes belonging to different clusters.
Also, since temporal link prediction is based on the time-evolving membership vector~$ \vphi $, 
it summarizes how accurately the learned cluster memberships capture temporally-evolving community structure.
\Cref{tab:results:linkpred} reports the link prediction accuracy in terms of 
the area under the receiver operating characteristic curve (AUC) and the average precision (AP).
Both metrics range from 0 to 1, and higher values are better.
As the number of test edges (\ie, $ E_{\tau'} \cup E_{\tau'}^{-} $) changes over time, 
we average the performance for each snapshot, weighted by the size of $ E_{\tau'} \cup E_{\tau'}^{-} $.
Results show that \method consistently outperforms baselines on all datasets, achieving up to 29\% higher temporal link prediction performance.
{The best results among baselines were mainly obtained by CTGCN, 
which is a temporal method that models the network evolution.
Among static baselines, AGCN mostly outperforms other statc methods, and even most dynamic baselines, except CTGCN.
This can be explained by the fact that these dynamic baselines are trained using cluster agnostic objectives,
which again shows that incorporating the clustering objective can be helpful for detecting communities.}
As~in \Cref{sec:exp:nodeclustering:temporal}, we report results obtained in the two settings (\ie, all vs. latest) for static baselines.
There is no clear winner between~them. \Cref{fig:performance_over_time:link_prediction} shows how the performance of \method and four baselines changes over time.

\begin{table}[!htp]\centering
\par\vspace{-1.2em}\par
\setlength{\abovecaptionskip}{0.0em}
\caption{\method consistently outperforms baselines, achieving up to 29\% higher temporal link prediction performance than the best baseline. 
Best results are in bold, and second best results are underlined.}\label{tab:results:linkpred}
\small
\makebox[0.4\textwidth][c]{
	\setlength{\tabcolsep}{0.4mm}

	\begin{tabular}{lrrrrrr}\toprule
		\multirow{2}{*}{Method} &\multicolumn{2}{c}{Foursquare-NYC} &\multicolumn{2}{c}{Foursquare-TKY} &\multicolumn{2}{c}{Yahoo-Msg} \\
		\cmidrule(l{1.5pt}r{1.5pt}){2-3} \cmidrule(l{1.5pt}r{1.5pt}){4-5} \cmidrule(l{1.5pt}r{1.5pt}){6-7}
&ROC AUC &Avg. Prec. &ROC AUC &Avg. Prec. &ROC AUC &Avg. Prec. \\\midrule
		SVD~\cite{golub1971singular} &9.68$\pm$0.3 &33.28$\pm$0.2 &4.18$\pm$0.0 &37.84$\pm$0.0 &59.51$\pm$0.5 &64.88$\pm$0.4 \\
		SVD-latest &17.67$\pm$0.6 &37.93$\pm$0.6 &7.20$\pm$0.2 &35.08$\pm$0.1 &49.26$\pm$0.2 &53.21$\pm$0.2 \\
		DGI~\cite{DBLP:conf/iclr/VelickovicFHLBH19} &14.37$\pm$0.7 &33.02$\pm$0.1 &13.79$\pm$1.0 &33.17$\pm$0.3 &50.60$\pm$0.5 &51.83$\pm$0.3 \\
		DGI-latest &18.55$\pm$1.2 &34.16$\pm$0.3 &20.01$\pm$0.7 &34.69$\pm$0.3 &41.92$\pm$0.2 &45.00$\pm$0.2 \\
		GAE~\cite{DBLP:journals/corr/KipfW16a} &13.55$\pm$1.1 &33.16$\pm$0.3 &17.44$\pm$0.5 &35.01$\pm$0.3 &46.40$\pm$0.5 &48.41$\pm$0.2 \\
		GAE-latest &19.80$\pm$0.4 &35.13$\pm$0.3 &21.67$\pm$0.7 &37.44$\pm$0.5 &42.45$\pm$0.5 &44.87$\pm$0.2 \\
		VGAE~\cite{DBLP:journals/corr/KipfW16a} &6.63$\pm$0.1 &32.34$\pm$0.1 &10.06$\pm$0.3 &34.90$\pm$0.4 &39.97$\pm$0.0 &47.99$\pm$0.1 \\
		VGAE-latest &12.02$\pm$0.2 &33.18$\pm$0.0 &12.91$\pm$0.2 &34.93$\pm$0.2 &44.21$\pm$0.1 &49.61$\pm$0.0 \\
		ARGA~\cite{DBLP:journals/tcyb/PanHFLJZ20} &6.96$\pm$0.0 &31.63$\pm$0.0 &11.45$\pm$0.1 &33.00$\pm$0.3 &38.79$\pm$0.1 &44.17$\pm$0.1 \\
		ARGA-latest &11.89$\pm$1.1 &32.38$\pm$0.2 &13.17$\pm$0.2 &32.61$\pm$0.0 &39.84$\pm$0.1 &43.78$\pm$0.0 \\
		ARGVA~\cite{DBLP:journals/tcyb/PanHFLJZ20} &13.56$\pm$0.4 &34.95$\pm$0.2 &22.30$\pm$0.4 &43.11$\pm$0.3 &46.99$\pm$0.1 &50.44$\pm$0.1 \\
		ARGVA-latest &26.01$\pm$0.7 &39.11$\pm$0.3 &32.01$\pm$0.5 &45.14$\pm$0.1 &50.54$\pm$0.1 &51.25$\pm$0.1 \\
		SDCN~\cite{DBLP:conf/www/Bo0SZL020} &47.86$\pm$0.7 &46.31$\pm$0.6 &37.32$\pm$0.8 &40.73$\pm$0.6 &55.76$\pm$1.5 &55.78$\pm$1.3 \\
		SDCN-latest &25.24$\pm$0.3 &36.47$\pm$0.3 &19.01$\pm$1.3 &35.05$\pm$0.9 &54.51$\pm$0.6 &55.35$\pm$0.5 \\
		AGCN~\cite{DBLP:conf/mm/PengLJH21} &\ul{56.13$\pm$1.0} &52.24$\pm$1.5 &42.43$\pm$2.7 &44.24$\pm$2.2 &54.23$\pm$2.2 &54.43$\pm$1.5 \\
		AGCN-latest &41.24$\pm$3.2 &49.01$\pm$2.5 &41.44$\pm$5.8 &51.27$\pm$4.0 &51.81$\pm$1.1 &52.87$\pm$0.4 \\ \midrule
		CTDNE~\cite{DBLP:conf/www/NguyenLRAKK18} &7.06$\pm$0.0 &31.55$\pm$0.0 &16.97$\pm$0.3 &33.59$\pm$0.1 &54.73$\pm$0.1 &54.16$\pm$0.1 \\
		CTDNE-latest &7.27$\pm$0.0 &32.28$\pm$0.0 &7.36$\pm$0.1 &31.98$\pm$0.0 &50.11$\pm$0.0 &52.70$\pm$0.1 \\
		TIMERS~\cite{DBLP:conf/aaai/ZhangCPW018} &23.84$\pm$0.2 &37.02$\pm$0.1 &15.09$\pm$0.1 &33.72$\pm$0.0 &48.87$\pm$0.1 &49.65$\pm$0.1 \\
		DynGEM~\cite{DBLP:journals/corr/abs-1805-11273} &26.65$\pm$0.8 &36.61$\pm$0.3 &25.52$\pm$2.8 &36.24$\pm$0.9 &47.46$\pm$0.5 &46.69$\pm$0.4 \\
		DynAERNN~\cite{DBLP:journals/kbs/GoyalCC20} &26.17$\pm$2.1 &41.39$\pm$1.6 &18.23$\pm$1.1 &40.15$\pm$0.7 &44.81$\pm$2.0 &50.44$\pm$2.1 \\
		EvolveGCN~\cite{DBLP:conf/aaai/ParejaDCMSKKSL20} &23.79$\pm$1.0 &47.45$\pm$0.1 &24.67$\pm$0.6 &46.45$\pm$0.2 &47.00$\pm$0.9 &47.08$\pm$0.4 \\
CTGCN~\cite{CTGCN} &50.58$\pm$2.4 &\ul{54.54$\pm$1.5} &\ul{51.61$\pm$4.5} &\ul{57.56$\pm$2.8} &\ul{75.51$\pm$0.9} &\ul{76.82$\pm$0.7} \\ \midrule\midrule
		\method (Ours) &\textbf{64.60$\pm$0.6} &\textbf{70.34$\pm$0.5} &\textbf{66.26$\pm$0.8} &\textbf{70.22$\pm$0.6} &\textbf{84.30$\pm$0.1} &\textbf{86.88$\pm$0.1} \\
		\bottomrule
	\end{tabular}

}
\vspace{-2.2em}
\end{table}

\vspace{-1.0em}
\subsection{Ablation Study (RQ3)}\label{sec:exp:ablation}
We investigate how contrastive learning objectives affects \method.
\Cref{fig:results:ablation} shows node clustering results where \method was trained with different combinations of contrastive objectives;
F, H, and C denote the loss terms on node features ($ \lambda_F $), 
network homophily ($ \lambda_H $), and hierarchical communities ($ \lambda_C $) in \cshref{eq:loss:combined}, respectively, and
only the specified terms were included with a weight of 1.
We report relative scores, \ie, scores divided by the best score for each metric.
Results show that the proposed contrastive objectives are complementary, 
\ie, jointly optimizing these objectives improves the performance, \eg,
F to F+H on \acm and H to H+C on \dblpS.
Especially, the best result on \acm and \dblpS are obtained when all objectives are used together (F+H+C).
However, \dblpS shows a different pattern, where the best result was obtained with F+C.
Notably, in \dblpS, the objective on network homophily was not useful whether it is used alone (H) or with others (F vs. F+H).
In \dblpS, 36\% of the nodes are isolated, making it hard to learn from graph structure.
Still, joint optimization improved the results \mbox{(\eg, H vs. H+C)}.

%% file: 060relatedwork.tex
\vspace{-0.7em}
\section{Related Work}\label{sec:relatedwork}

\textbf{Graph Clustering.}
Several approaches have been developed or adapted for graph clustering and community detection,
including modularity-based methods~\cite{girvan2002community}, METIS~\cite{DBLP:journals/siamsc/KarypisK98}, 
spectral methods~\cite{DBLP:conf/ppsc/BarnardS93}, methods based on SVD~\cite{golub1971singular}, 
connected components~\cite{DBLP:conf/icdm/ParkPMK16,10.1371/journal.pone.0229936},
tensor factorization~\cite{DBLP:conf/www/GujralPP20,DBLP:journals/vldb/ParkOK19,DBLP:conf/icde/ParkOK17,DBLP:conf/icde/OhPSK18}
and MDL (Minimum Description Length)~\cite{DBLP:conf/sdm/AkogluTMF12,DBLP:conf/kdd/SunFPY07}.
However, these methods all miss one or more of the desiderata of \Cref{tab:salesmatrix},
as they mostly focus on utilizing the graph structure alone, 
with no support for input node features or the time evolution of graphs, and 
without learning node representations, which are useful for downstream applications.
Our comparison with SVD \cite{golub1971singular}, one of the representative methods for community detection,
shows the benefits of satisfying the desiderata in \Cref{tab:salesmatrix}.

In this paper, we focus on another group of methods for graph clustering, 
namely, deep graph clustering (DGC).
Methods for DGC can be grouped into two categories: 
(1) two-stage methods that perform clustering after learning representations, and 
(2) single-stage methods that jointly perform clustering and representation learning (RL).
Unsupervised graph RL methods are used for two-stage deep graph clustering (DGC).
In~\cite{DBLP:conf/aaai/TianGCCL14}, for instance, AEs are used to learn non-linear node embeddings, 
and then K-means is applied to get clustering assignments.
GNN-based encoders are adopted in more recent methods.
GAE~\cite{DBLP:journals/corr/KipfW16a} and VGAE~\cite{DBLP:journals/corr/KipfW16a} 
learn node embeddings using a graph autoencoder and a variational variant.
ARGA~\cite{DBLP:journals/tcyb/PanHFLJZ20} and ARGVA~\cite{DBLP:journals/tcyb/PanHFLJZ20} employ
an adversarially regularized graph autoencoder and its variational version.
A few recent studies~\cite{DBLP:conf/iclr/VelickovicFHLBH19,DBLP:conf/icml/YouCWS20,DBLP:conf/kdd/WangLHS21,DBLP:conf/aaai/SunLZ20} 
investigated self-supervised learning techniques for graph RL,
\eg, DGI~\cite{DBLP:conf/iclr/VelickovicFHLBH19} optimizes GCN encoder 
by contrasting node embeddings with the embedding of the graph.

DMoN~\cite{tsitsulin2020graph} is a single-stage method that performs clustering via spectral modularity maximization.
DAEGC~\cite{DBLP:conf/ijcai/WangPHLJZ19} simultaneously optimizes embedding learning and graph clustering
by combining the clustering loss of DEC with the graph reconstruction loss of graph attentional AEs.
SDCN~\cite{DBLP:conf/www/Bo0SZL020} improves DAEGC by integrating a GCN encoder and AEs via a delivery operator.
AGCN~\cite{DBLP:conf/mm/PengLJH21} further improves upon SDCN by developing two attention-based fusion modules, 
which aggregate features from GCNs and AEs, and multi-scale features from different layers.
Despite some differences (\eg, encoder architectures),
existing DGC methods are mainly based on AEs, involve reconstruction loss minimization, and 
use the same clustering objective~\cite{DBLP:conf/icml/XieGF16} with small adjustments.
The proposed \method performs deep graph clustering
in a novel contrastive graph learning framework with multi-level contrastive objectives.

\textbf{Temporal Graph Clustering (TGC).}
Existing methods mainly perform TGC based on the graph structure and its temporal change, 
without considering node features and their semantics in the clustering objective.
Existing TGC methods can be grouped into two classes:
snapshot clustering~\cite{DBLP:conf/kdd/ChiSZHT07,DBLP:conf/kdd/Berger-WolfS06,DBLP:conf/asunam/GreeneDC10} and consensus clustering~\cite{lancichinetti2012consensus,rosvall2008maps,rosvall2011multilevel,aynaud2011multi,10.1371/journal.pone.0195993}.
Given graph snapshots, each snapshot is clustered separately in snapshot clustering, thereby ignoring inter-snapshot information.
Consensus clustering instead finds a single partitioning for the entire graph snapshots.
Consensus and snapshot clustering correspond to two fixed choices (\ie, the entire snapshots vs. the last one), which is not always optimal.
\method instead adaptively determines a subset of snapshots to find clusters from.

For two-stage deep TGC, unsupervised dynamic graph representation learning methods can also be employed,
which learn dynamic embeddings using temporal random walk~\cite{DBLP:conf/www/NguyenLRAKK18}, incremental SVD~\cite{DBLP:conf/aaai/ZhangCPW018}, 
AEs~\cite{DBLP:journals/corr/abs-1805-11273,DBLP:journals/kbs/GoyalCC20}, and
RNNs combined with GCNs~\cite{DBLP:conf/aaai/ParejaDCMSKKSL20,CTGCN,DBLP:conf/wsdm/evokg}.
Yet no single-stage DGC methods have been designed for TGC.
This paper presents the first such method for temporal network analysis.

\begin{figure}[!t]
\par\vspace{-1.2em}\par
\centering
\makebox[0.4\textwidth][c]{
	\includegraphics[width=0.90\linewidth]{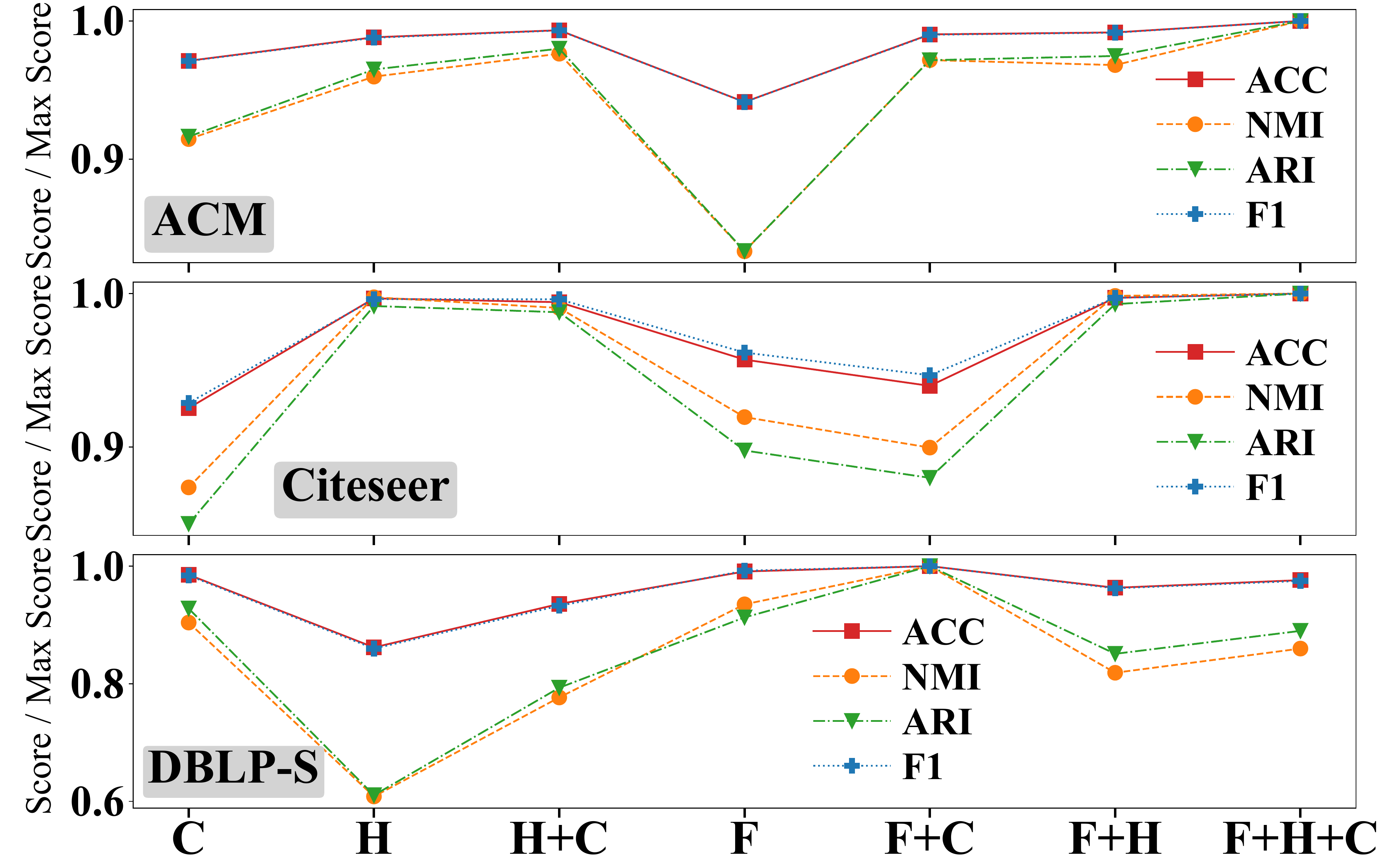}
}
\setlength{\abovecaptionskip}{0.0em}
\caption{
Node clustering performance of \method, obtained with different contrastive objectives.
F: Node Features. H: Network Homophily. C: Hierarchical Communities.
}
\label{fig:results:ablation}
\vspace{-2.2em}
\end{figure}

%% file: 070conclusion.tex
\vspace{-0.8em}
\section{Conclusion} \label{sec:conclusion}
\vspace{-0.2em}

This work presented \method, a new deep graph clustering framework
for community detection and tracking in the web data.
\begin{itemize}[leftmargin=*,topsep=0pt]
\item \contribFramework
\method jointly learns node embeddings and cluster memberships in a novel contrastive graph learning framework.
\method effectively finds clusters by using information along 
multiple dimensions, \eg, node features, hierarchical \mbox{communities}.

\item \contribTemporal
\method is designed to find clusters from time-evolving graphs, improving upon existing deep graph clustering methods, which are designed for static graphs.

\item \contribEffective
We show the effectiveness of \method via extensive evaluation on several static and temporal real-world graphs.
\end{itemize}

%% file: 080appendix.tex
\section{Mining Case Studies}\label{sec:app:casestudy}

\subsection{Case Studies on Synthetic Graphs}

In this section, we show how effectively \method performs community detection and tracking, 
using synthetic graphs that consist of a small number of groups;
each group corresponds to a tightly knit community, which experiences significant changes over time.

\begin{figure}[h!]
	\par\vspace{-0.5em}\par
	\centering
	\makebox[0.4\textwidth][c]{
		\begin{subfigure}[t]{0.12\textwidth}
			\centering
			\includegraphics[width=\textwidth]{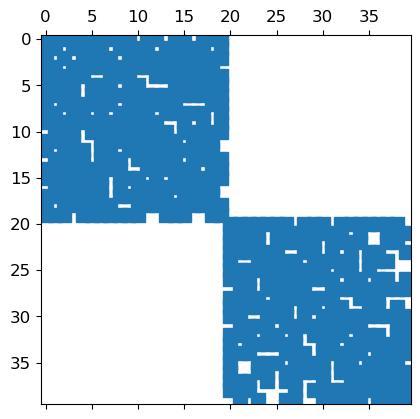}
			\setlength{\abovecaptionskip}{-1.0em}
			\caption{Segment before the change point (time 3)}
			\label{fig:synth1:prevseg:all}
		\end{subfigure}
\begin{subfigure}[t]{0.12\textwidth}
			\centering
			\includegraphics[width=\textwidth]{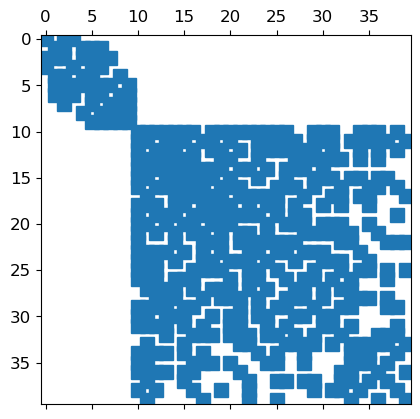}
			\setlength{\abovecaptionskip}{-1.0em}
			\caption{Segment after segmentation at the change point}
			\label{fig:synth1:curseg:all:withseg}
		\end{subfigure}
\begin{subfigure}[t]{0.12\textwidth}
			\centering
			\includegraphics[width=\textwidth]{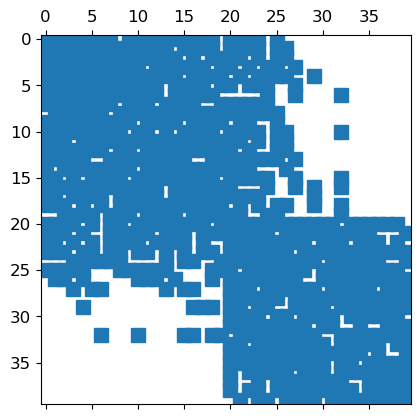}
			\setlength{\abovecaptionskip}{-1.0em}
			\caption{Segment with no segmentation at change point}
			\label{fig:synth1:curseg:all:withoutseg}
		\end{subfigure}
\begin{subfigure}[t]{0.15\textwidth}
			\centering
			\includegraphics[width=\textwidth]{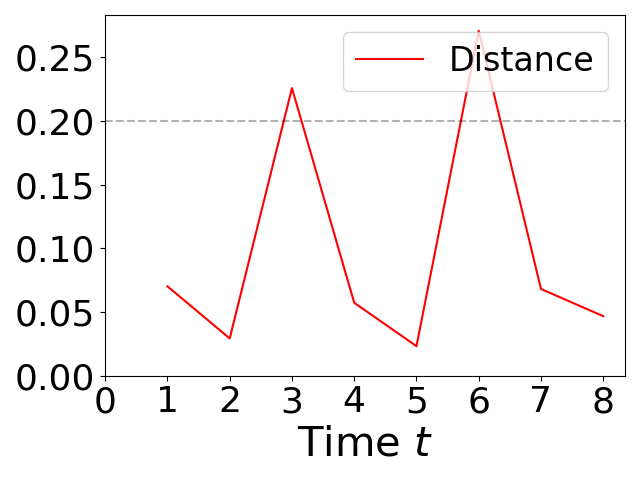}
			\setlength{\abovecaptionskip}{-1.0em}
			\caption{Distance between $ G_t $ and the existing segment over time}
			\label{fig:synth1:distance}
		\end{subfigure}
	}
	\captionsetup{width=1.025\linewidth}
	\setlength{\abovecaptionskip}{0.5em}
	\caption{Two groups with traveling members (Case 1). Segmentation reveals a clearer community structure across time.}
	\label{fig:synth1}
	\vspace{-1.0em}
\end{figure}

\noindent
\textbf{Case 1: Two Groups With Traveling Members} (\Cref{fig:synth1}).
We have groups 1 and 2 for time 0-2. 
At time 3, half of the nodes in group 1 move to group 2, stay there until time 5, and then at time~6,
move back to group 1, where they originally belonged.
Thus there are two change points (CPs), \ie, time 3 and 6 (\Cref{fig:synth1:distance}).
\Cref{fig:synth1:prevseg:all} shows the segment prior to the first CP.
\Cref{fig:synth1:curseg:all:withseg,fig:synth1:curseg:all:withoutseg} show
the segment at the first CP when the graph stream was properly segmented or not;
\Cref{fig:synth1:curseg:all:withoutseg} does not clearly show the change in the size of two groups.
By performing segmentation in the presence of a significant change, \method captures a clearer community structure.

\begin{figure}[h!]
	\par\vspace{-0.5em}\par
	\centering
	\makebox[0.4\textwidth][c]{
		\begin{subfigure}[t]{0.12\textwidth}
			\centering
			\includegraphics[width=\textwidth]{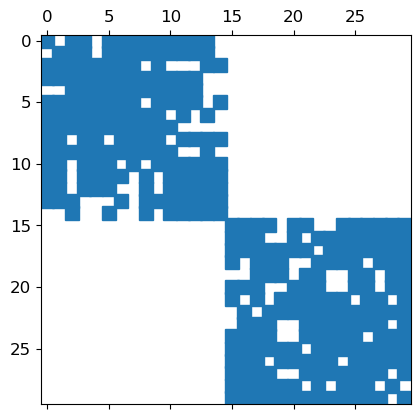}
			\setlength{\abovecaptionskip}{-1.0em}
			\caption{Segment before the change point (time 3)}
			\label{fig:synth2:prevseg:all}
		\end{subfigure}
\begin{subfigure}[t]{0.12\textwidth}
			\centering
			\includegraphics[width=\textwidth]{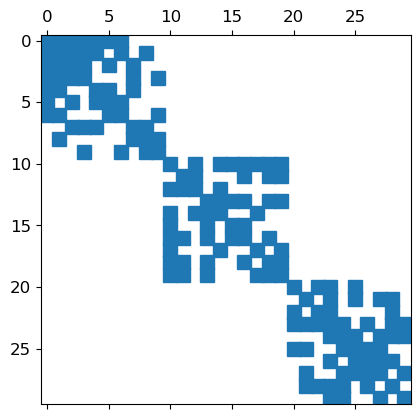}
			\setlength{\abovecaptionskip}{-1.0em}
			\caption{Segment after segmentation at the change point}
			\label{fig:synth2:curseg:all:withseg}
		\end{subfigure}
\begin{subfigure}[t]{0.12\textwidth}
			\centering
			\includegraphics[width=\textwidth]{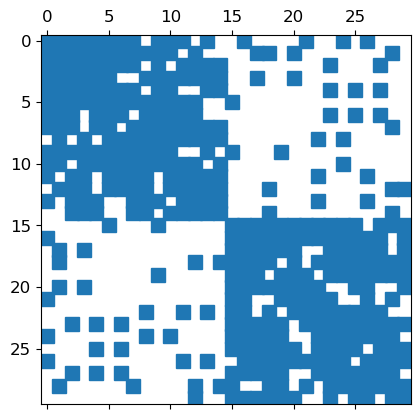}
			\setlength{\abovecaptionskip}{-1.0em}
			\caption{Segment with no segmentation at change point}
			\label{fig:synth2:curseg:all:withoutseg}
		\end{subfigure}
\begin{subfigure}[t]{0.15\textwidth}
			\centering
			\includegraphics[width=\textwidth]{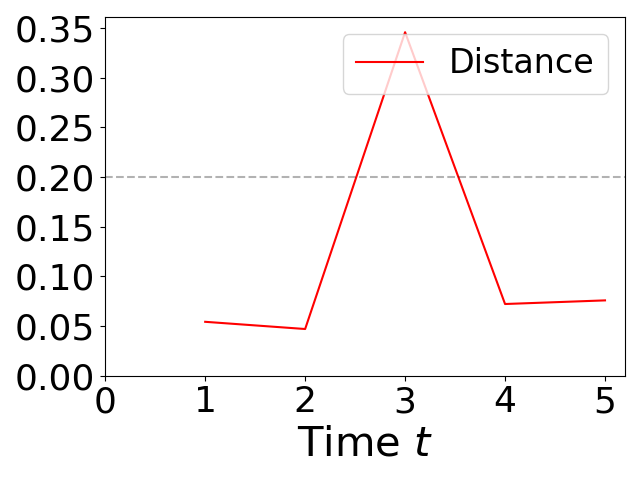}
			\setlength{\abovecaptionskip}{-1.0em}
			\caption{Distance between $ G_t $ and the existing segment over time}
			\label{fig:synth2:distance}
		\end{subfigure}
 	}
	\captionsetup{width=1.06\linewidth}
	\setlength{\abovecaptionskip}{0.5em}
	\caption{Two groups reorganizing into three (Case 2). 
		\method identifies reorganizing communities and detects the change~point.
	}
	\label{fig:synth2}
	\vspace{-1.0em}
\end{figure}

\noindent
\textbf{Case 2: Two Groups Reorganizing Into Three} (\Cref{fig:synth2}).
This network initially consists of two communities, which are regrouped into three communities due to a major reorganization at time~3.
\Cref{fig:synth2:prevseg:all} shows the two communities captured by \method before the CP at time 3.
\Cref{fig:synth2:curseg:all:withseg} shows that \method successfully detects the CP (\Cref{fig:synth2:distance}), and 
discovers restructured communities.
Again, when the CP is ignored, it gets harder to see a clear structure of three communities from the resulting graph stream segment  (\Cref{fig:synth2:curseg:all:withoutseg}).

\begin{figure}[t!]
	\par\vspace{-1.5em}\par
	\centering
	\begin{subfigure}[t]{0.30\textwidth}
		\centering
		\includegraphics[width=\textwidth]{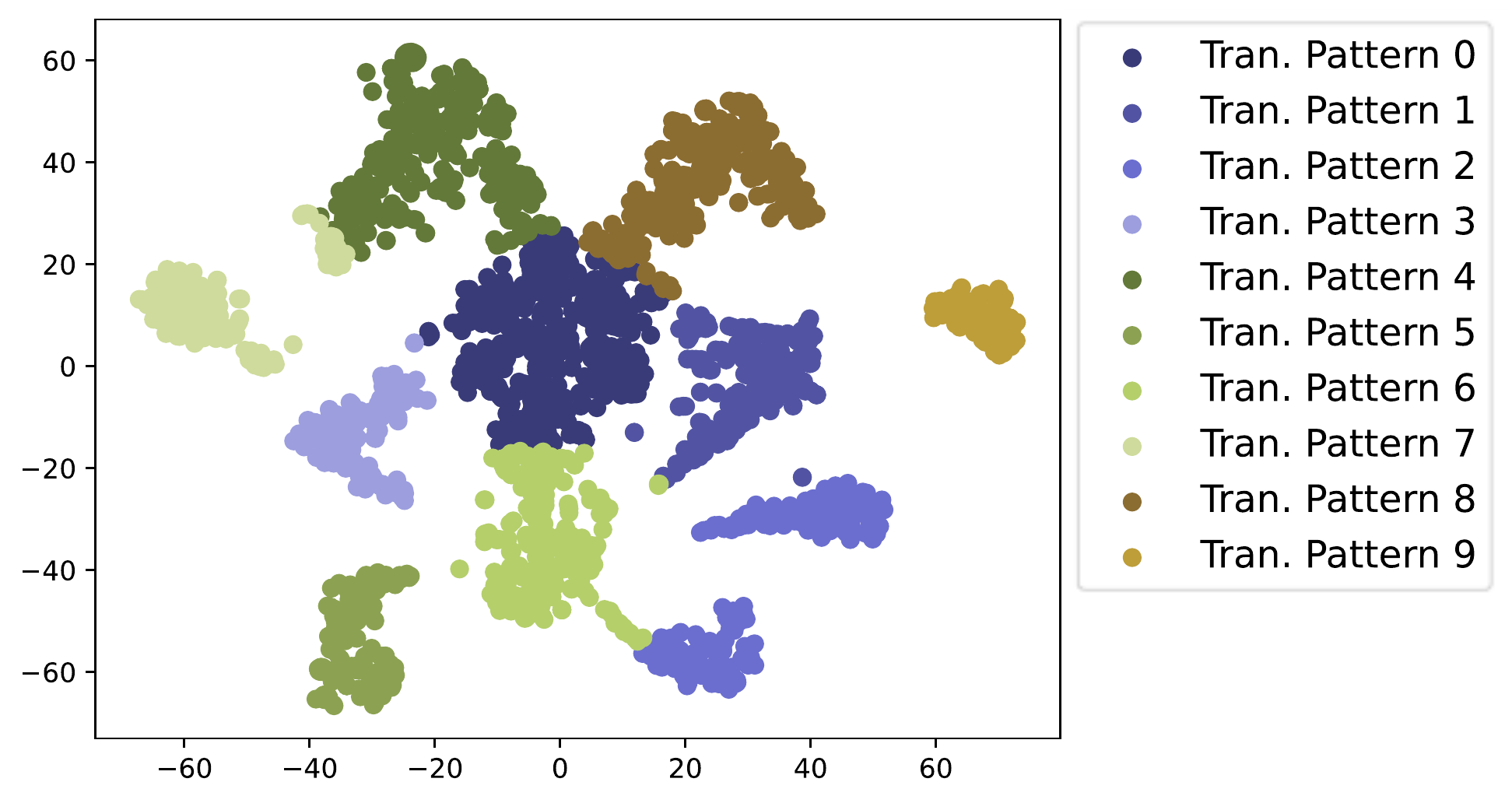}
\end{subfigure}
\begin{subfigure}[t]{0.21\textwidth}
		\centering
		\includegraphics[width=\textwidth]{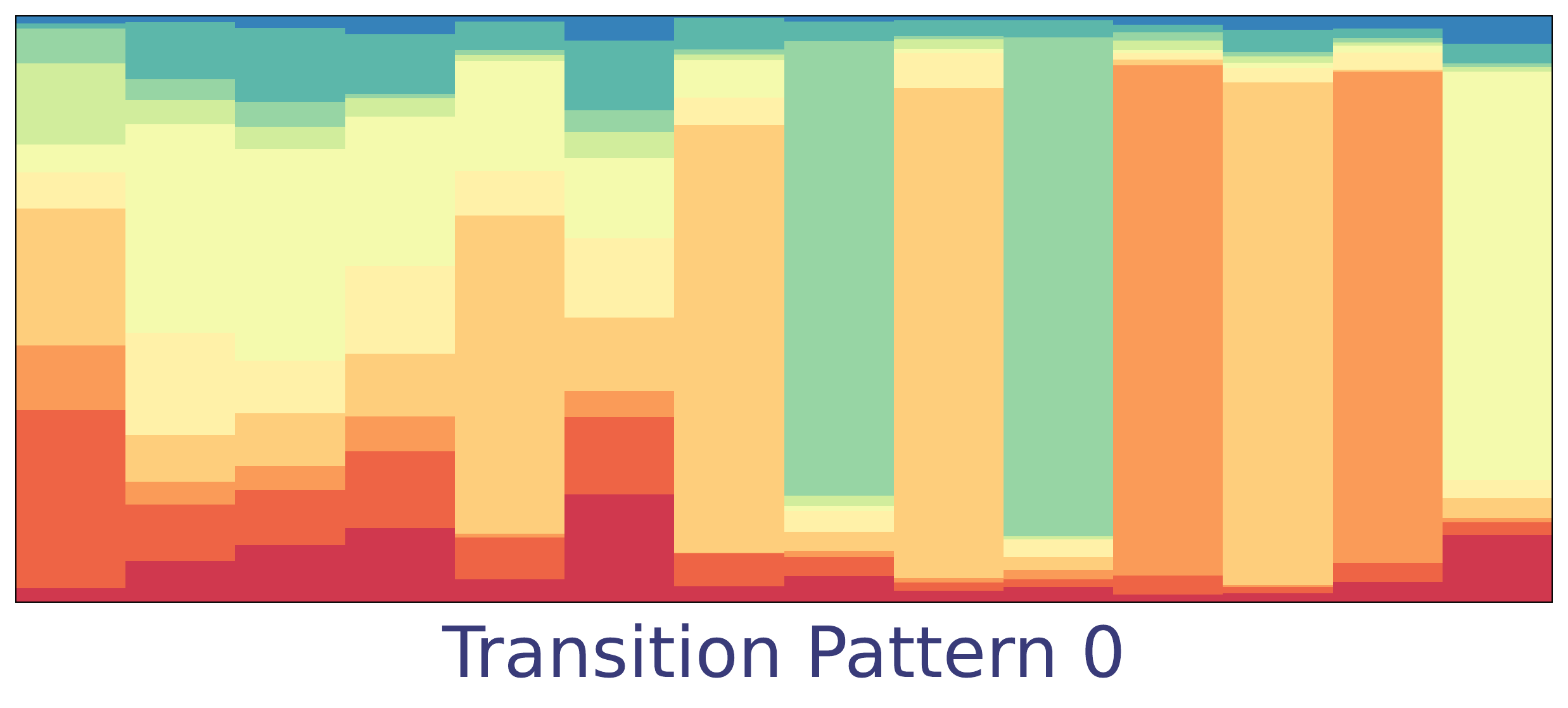}
\end{subfigure}
\begin{subfigure}[t]{0.21\textwidth}
		\centering
		\includegraphics[width=\textwidth]{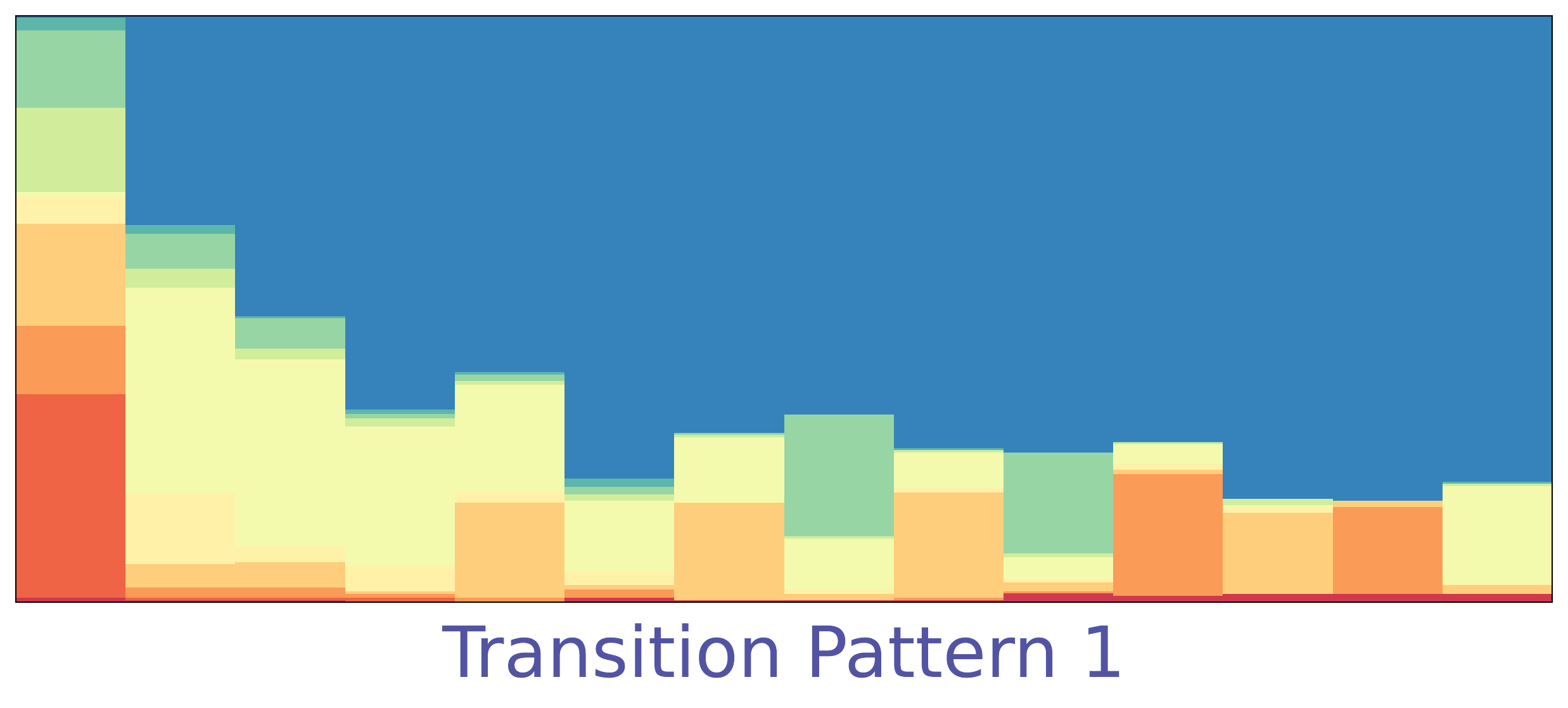}
\end{subfigure}
\begin{subfigure}[t]{0.21\textwidth}
		\centering
		\includegraphics[width=\textwidth]{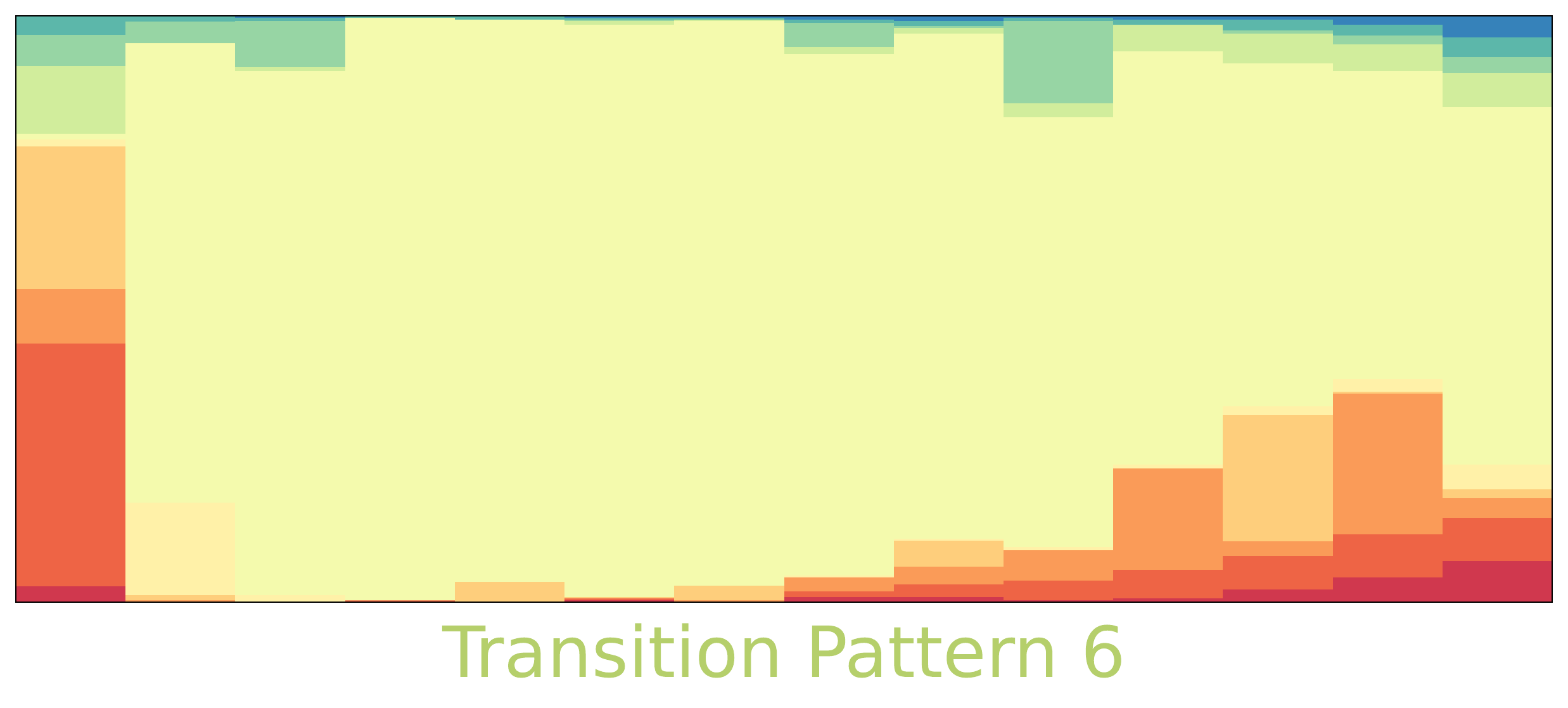}
\end{subfigure}
\begin{subfigure}[t]{0.21\textwidth}
		\centering
		\includegraphics[width=\textwidth]{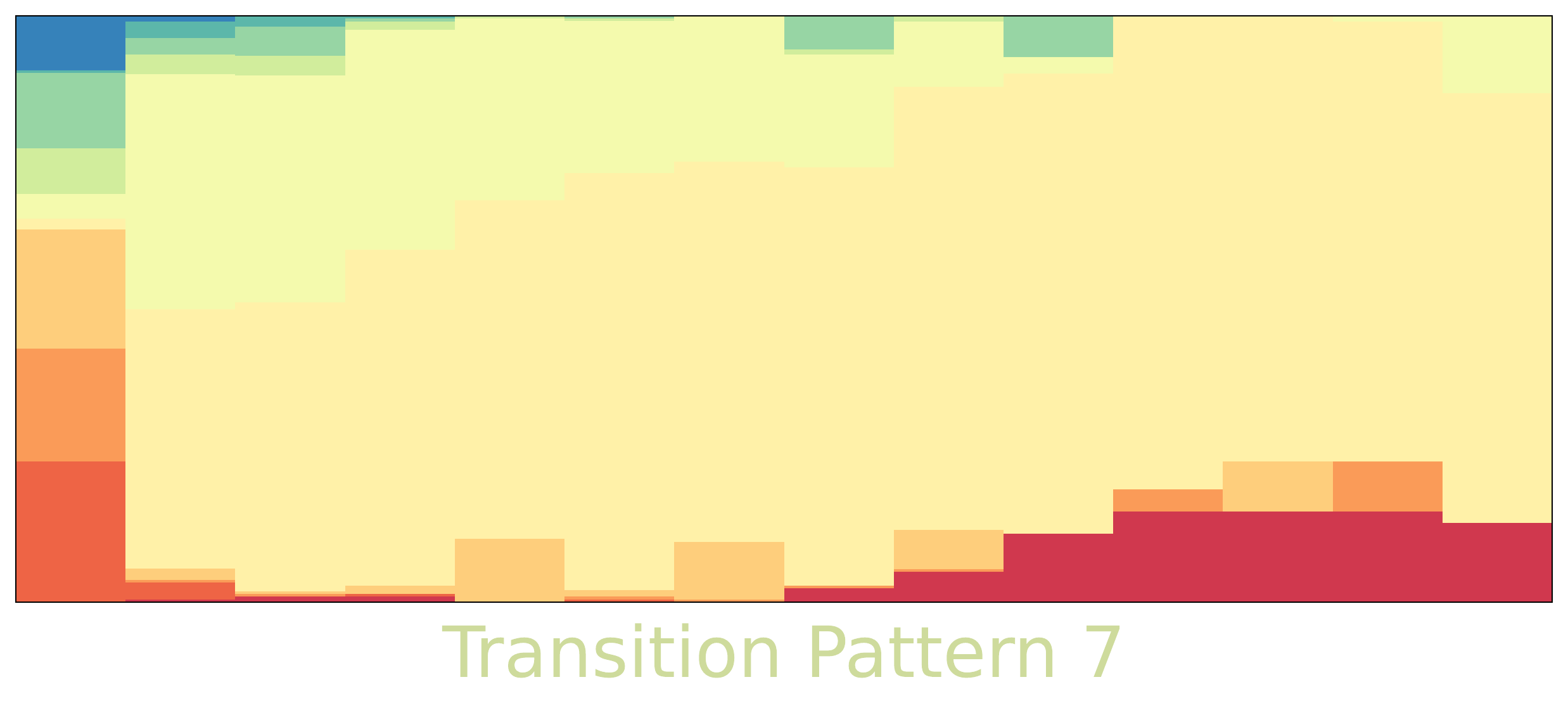}
\end{subfigure}
	\setlength{\abovecaptionskip}{0.0em}
	\caption{Node clusters (top) based on their transition patterns (bottom) in the \yahooMsg dataset.
}
	\label{fig:transpattern:yahoo}
	\vspace{-2.0em}
\end{figure}

\begin{figure*}[htbp!]
	\par\vspace{-1.5em}\par
	\centering
	\makebox[0.4\textwidth][c]{
		\begin{subfigure}[t]{0.42\textwidth}
			\centering
			\includegraphics[width=\textwidth]{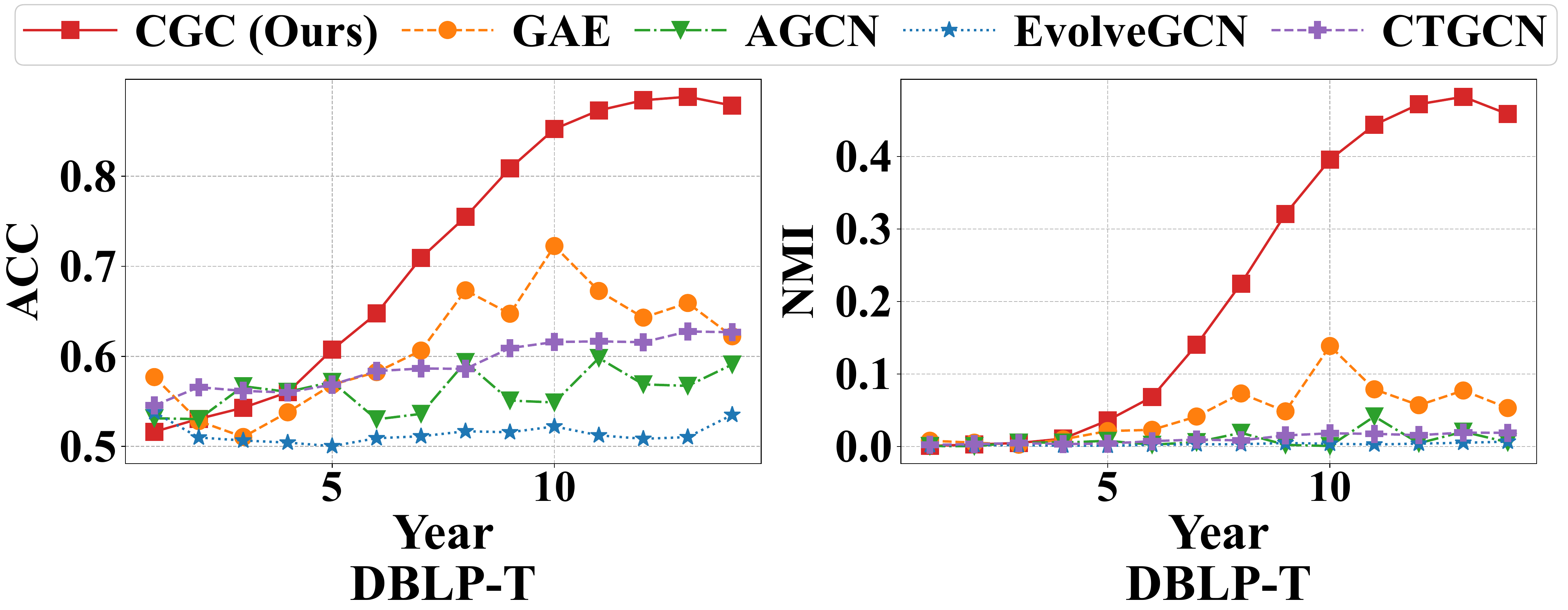}
			\setlength{\abovecaptionskip}{-1.0em}
			\caption{Temporal Node Clustering}
			\label{fig:performance_over_time:node_clustering}
		\end{subfigure}
		\begin{subfigure}[t]{0.56\textwidth}
			\centering
			\includegraphics[width=\textwidth]{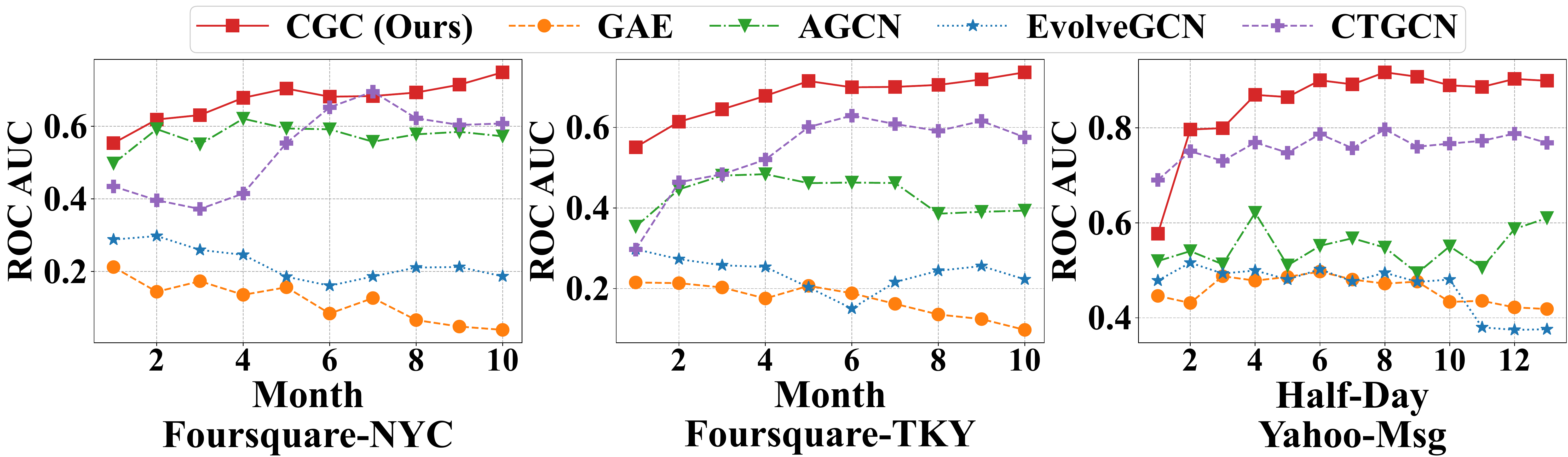}
			\setlength{\abovecaptionskip}{-1.0em}
			\caption{Temporal Link Prediction}
			\label{fig:performance_over_time:link_prediction}
		\end{subfigure}
	}
\setlength{\abovecaptionskip}{0.1em}
	\caption{\method achieves the best clustering performance nearly consistently on temporal graphs over the entire time period.}
	\label{fig:performance_over_time}
	\vspace{-1.0em}
\end{figure*}

\vspace{-0.8em}
\subsection{Case Studies on Real-World Graphs}
\vspace{-0.2em}

To see how the cluster membership found by \method evolves over time, 
we cluster nodes based on the transition pattern (TP) of their membership vectors (\Cref{fig:transpattern:yahoo} (top)).
Specifically, we concatenate the cluster membership vectors of each node obtained at different time steps,
apply t-SNE to embed nodes in a two-dimensional space, and 
perform $ k $-means clustering on the resulting two-dimensional node embeddings to obtain TP clusters.
Then for each TP, we consider how cluster distribution changed over time (\Cref{fig:transpattern:yahoo} (bottom)).
For each time step, we take the average of the membership vectors of the nodes belonging to a specific TP, and
display the cluster distribution at each time as a column;
clusters are associated with distinct colors, and 
the cluster distribution in the averaged membership vector at different time is shown by the proportion of the corresponding~colors.

\textbf{\yahooMsg (\Cref{fig:transpattern:yahoo}).}
Nodes are clustered into 10 TPs.
Among them, TP 0 shows a different pattern than others, 
where a major cluster changes frequently over time (e.g., switches between orange and green). 
In the scatter plot above, TP 0 is the cluster at the center, located close to a few surrounding clusters.
Over time, the cluster assignments of nearby clusters have had a varying impact on how the nodes in TP~0 are clustered.
Also, note that a segmentation occurred at the second time step, as can be seen in the TP plots.
The color distribution of the first column in the four TPs greatly differs from those of the second and subsequent columns.
Via segmentation, \method discovers a clearer community structure.

\vspace{-1.0em}
\section{Clustering Performance over Time}\label{sec:app:performance_over_time}
\vspace{-0.3em}

\Cref{fig:performance_over_time} shows how the performance of \method and four select baselines changes over time.
For static baselines, we report the results obtained by clustering all observed graph snapshots at each time~step.

\textbf{Node Clustering} (\Cref{fig:performance_over_time:node_clustering}). 
While all methods do not perform well for the first few time steps,
\method's performance continuously improves over time, reaching an ACC of $ \sim $0.89 and an NMI of $ \sim $0.48 in the end.
Although baselines' performance also improves with time, 
their improvement is much smaller than that of \method, failing to effectively track the evolution of communities in the network.

\textbf{Link Prediction} (\Cref{fig:performance_over_time:link_prediction}). 
\method significantly outperforms baselines throughout most of the time span.
Dynamic methods are not effective at capturing community structure, 
while deep clustering baselines like AGCN fail to track the evolution of clusters.

\begin{table*}[!t]\centering
\setlength{\abovecaptionskip}{0.0em}
	\caption{Summary of temporal real-world datasets. \textsc{n/a} denotes that the corresponding datasets do not have node labels.}
	\label{tab:datasets:temporal}
	\small
	\makebox[0.4\textwidth][c]{
		\renewcommand{\arraystretch}{0.90}
		\setlength{\tabcolsep}{1.4mm}
		\begin{tabular}{lrrrrrrrr}\toprule
			\textbf{\makecell[c]{Dataset}} &\textbf{\makecell[c]{Edge Type\\(node i, node j, time t)}} &\textbf{\makecell[c]{\# Nodes}} &\textbf{\makecell[c]{\# Edges}} &\textbf{\makecell[c]{Time Range\\(Inclusive)}} &\textbf{\makecell[c]{\# Graph\\Snapshots}} &\textbf{\makecell[c]{Snapshot\\Interval}} &\textbf{\makecell[c]{\# Dynamic\\Node Classes}} \\\midrule
			\yahooMsg &(user, user, time-second) &82,309 (82,309 users) &786,911 &0-6 (days) &14 &12 hours &\textsc{n/a} \\
			\foursquareNYC &(user, venue, time-second) &39,416 (1,083 users, 38,333 venues) &454,856 &0-318 (days) &11 &30 days &\textsc{n/a} \\
			\foursquareTKY &(user, venue, time-second) &64,151 (2,293 users, 61,858 venues) &1,147,406 &0-318 (days) &11 &30 days &\textsc{n/a} \\
			\dblpT &(author, author, time-year) &6,942 (6,942 authors) &168,124 &0-13 (years) &14 &1 year &2 \\
			\bottomrule
		\end{tabular}
	}
	\vspace{-1.5em}
\end{table*}

\begin{table}[!t]\centering
	\setlength{\abovecaptionskip}{0.0em}
	\caption{Summary of static real-world datasets used in experiments. In all datasets, nodes have labels and input features.}
	\label{tab:datasets:static}
	\small
	\makebox[0.4\textwidth][c]{
		\setlength{\tabcolsep}{0.8mm}
		\renewcommand{\arraystretch}{0.90}
		\begin{tabular}{lrrrrr}\toprule
			\textbf{\makecell[c]{Dataset}} &\textbf{\makecell[c]{Edge Type\\(node i, node j)}} &\textbf{\makecell[c]{\# Nodes}} &\textbf{\makecell[c]{\# Edges}} &\textbf{\makecell[c]{\# Node\\Classes}} &\textbf{\makecell[c]{Feature\\Dimension}} \\\midrule
			\acm &(paper, paper) &3,025 &26,256 &3 &1,870 \\
			\dblpS &(author, author) &4,057 &7,056 &4 &334 \\
			\citeseer &(document, document) &3,327 &9,104 &6 &3,703 \\
			\magCS &(author, author) &18,333 &163,788 &15 &6,805 \\
			\bottomrule
		\end{tabular}
}
	\vspace{-2.5em}
\end{table}

\vspace{-0.5em}
\section{Experimental Settings}\label{sec:app:settings}
\vspace{-0.2em}

\hspace*{0.8\parindent}
\textbf{Experiments for Static Data.}
For \acm, \dblpS, and \citeseer, we cite the results of all baselines (except SVD, DGI and AGCN) from \cite{DBLP:conf/www/Bo0SZL020}.
For AGCN, we take its result from \cite{DBLP:conf/mm/PengLJH21}.
Settings of these baselines are given in~\cite{DBLP:conf/www/Bo0SZL020,DBLP:conf/mm/PengLJH21}.
We directly evaluate SVD and DGI on these datasets.
On \magCS, we evaluate all baselines using the settings in~\cite{DBLP:conf/www/Bo0SZL020,DBLP:conf/mm/PengLJH21}.
For methods we evaluate, we report results averaged over 5 runs.
We set node embedding size to 200 for SVD~\cite{sklearn:code} and DGI~\cite{DGI:code}.
We use a single-layer GCN in DGI as in the open source code~\cite{DGI:code}.
For \method, we set node embedding size to 200,~and~used Adam optimizer with a weight decay of 0.0001.
We set the learning rate to 0.0005 (\citeseer), 0.001 (\acm, \magCS), and 0.005 (\dblpS).
We used a single layer GNN in \method. 
We set temperature $ \tau $ to 0.65, $ \delta $ to $ 0.7 $, and update interval $ R = 2$ in all experiments.
Let $ r_F, r_H $, and $r^{\ell}_C $ be the number of negatives per positive sample 
for the contrastive loss $ \mathcal{L}_F $, $ \mathcal{L}_H $, and $ \mathcal{L}_C $, 
where $ \ell $ in $ r^{\ell}_C $ refers to the $ \ell $-th level clusters.
We set $ r_F $ to 180 (\magCS), 50 (\dblpS), and 30 (\acm, \citeseer);
$ r_H $ to 60 (\magCS) and 10 (others);
$ r^{\ell}_C $ to 60 (\magCS) and 30 (others) for each $ \ell $.
Let $ k $ denote the number of clusters to find. We set $ \set{K}\!=\!\{ k, 5k, 25k \} $.
For \dblpS, we set $ \lambda_F\!=\!4 $, $ \lambda_H\!=\!0 $, $ \lambda_C\!=\!1 $.
For \acm, \citeseer, and \magCS, we set $ \lambda_F\!=\!1 $,~$ \lambda_H\!=\!1 $,~$ \lambda_C\!=\!1 $.

\textbf{Experiments for Temporal Data.} 
Since the temporal graphs used in experiments have no input node features $ \mF $, 
we used learnable node embeddings as the input node features,
which were initialized by applying SVD to the row normalized adjacency matrix.

For both node clustering (\Cref{tab:results:nodeclus:temporal}) and 
link prediction (\Cref{tab:results:linkpred}) evaluation, 
baselines used mostly the same settings.
We set the size of initial node features and latent node embeddings to 128 and 32, respectively, and
used the Adam optimizer with a learning rate of 0.001.
Since the datasets used for temporal link prediction (\yahooMsg, \foursquareNYC, \foursquareTKY) do not have ground truth clusters,
we set the size of cluster membership to 64 for all baselines and \method.
\Cref{tab:results:nodeclus:temporal,tab:results:linkpred} report results averaged over five runs.

For SVD and DGI, we used the same setting used for static graphs.
For GAE, VGAE, ARGA, and ARGVA, we used the implementation of the PyTorch Geometric~\cite{PYG:code} with two-layer GCN encoders.
For SDCN and AGCN, we used the default settings used in \cite{DBLP:conf/www/Bo0SZL020,DBLP:conf/mm/PengLJH21}, 
while setting the size of node embeddings to 32.
For CTDNE, we used the default settings of the open source implementation~\cite{CTDNE:code}.
We set $ \theta $ in TIMERS to 0.17.
In DynGEM, we set $ \alpha $ to $ 10^{-5} $, $ \beta $ to $ 10 $, and both $ \nu_1 $ and $ \nu_2 $ to $ 10^{-4} $.
For DynAERNN, we set $ \beta $ to $ 5 $, the look back parameter to $ 3 $, and both $ \nu_1 $ and $ \nu_2 $ to $ 10^{-6} $.
In EvolveGCN, we used a two-layer GCRN; specifically, we used EvolveGCN-H, which incorporates node embeddings in RNNs.
For CTGCN, we used the CTGCN-C version with the settings used in \cite{CTGCN}.
In \method, we set $ \lambda_H\!=\!1 $, $ \lambda_C\!=\lambda_T\!=\!0.2 $, $ \lambda_F\!=\!0 $;
$ \psi\!=\!0.99 $, $ \theta\!=0.3 $.
Let $ r_T $ be the number of negatives per positive sample for the loss $ \mathcal{L}_T $.
For all temporal datasets, we set $ r_F\!=\!r_H\!=\!r_T\!=\!10 $.
We set $ r^{\ell}_C $ to 60 (link prediction datasets) and 30 (\dblpT) for each $ \ell $.
We set $ \set{K}\!=\!\{ 5k, 25k \} $ for \dblpT, and $ \set{K}\!=\!\{ k, 5k, 25k \} $ for all others.
For \method, we set the learning rate to 0.005, and the node embedding size to 32.

\vspace{-0.8em}
\section{Graph Stream Segmentation}\label{sec:app:segmentation}
\vspace{-0.4em}

\cshref{alg:segmentation} shows how \method decides whether to segment the graph stream or not.
A description of \cshref{alg:segmentation} is given in \cshref{sec:framework:temporal:segmentation}.

{\renewcommand{\setAlgFontSize}{\small}\renewcommand{\multilinenospace}[1]{\State \parbox[t]{\dimexpr\linewidth-\algorithmicindent}{\begin{spacing}{1.0}\setAlgFontSize#1\strut \end{spacing}}}
\renewcommand{\multilinenospaceD}[1]{\State \parbox[t]{\dimexpr0.96 \linewidth-\algorithmicindent}{\begin{spacing}{1.0}\setAlgFontSize#1\strut \end{spacing}}}

\algblockdefx[parallel]{parfor}{endpar}[1][]{$\textbf{parallel for}$ #1 $\textbf{do}$}{$\textbf{end parallel}$}
\algrenewcommand{\alglinenumber}[1]{\fontsize{7.5}{8}\selectfont#1\;\;}
\begin{figure}[h!]
	\par\vspace{-1.0em}\par
	\vspace{-5mm}
	\begin{center}
		\begin{algorithm}[H]
			\caption{\,\textsf{GraphStreamSegmentation}}
			\label{alg:segmentation}
			\begin{algorithmic}[1]
				\Require graph stream segment $\mathcal{G}_{\text{seg}}$, 
				new graph $G_{\tau_{j+1}}$ for time span $j\!+\!1$,
				input node features $\mF \in \RR^{\numNodes \times \featDim}$, 
segmentation threshold~$ \theta $
\Ensure 
				graph stream segment $ \mathcal{G}_{\text{seg}} $

				\If{$ \mathcal{G}_{\text{seg}} \ne \varnothing $}
				\State $ G_{\text{seg}} = \textsf{Merge}(\mathcal{G}_{\text{seg}}) $
				\State $ V^* = \textsf{Nodes}(G_{\text{seg}}) \cap \textsf{Nodes}(G_{\tau_{j+1}}) $
				\State $ \mH^{\text{seg}} = \mathcal{E}(G_{\text{seg}}, \mF ) $ 
				\State $ \mH^{j+1} = \mathcal{E}(G_{\tau_{j+1}}, \mF ) $ 
				\EndIf
				
				\If{$ \mathcal{G}_{\text{seg}} = \varnothing \textbf{~or~} \textsf{Dist}( \mH^{\text{seg}}_{V^*}, \mH^{j+1}_{V^*} ) > \theta $}
				\State $ \mathcal{G}_{\text{seg}} = \{ G_{\tau_{j+1}} \}$ \Comment{Start a new graph stream segment.}
				\Else
				\State $ \mathcal{G}_{\text{seg}} = \mathcal{G}_{\text{seg}} \cup \{ G_{\tau_{j+1}} \} $ \Comment{Add $G_{\tau_{j+1}}$ to the current segment.}
				\EndIf

				\State \Return $ \mathcal{G}_{\text{seg}} $
				
				\smallskip
			\end{algorithmic}
		\end{algorithm}
		\vspace{-3em}
	\end{center}
\end{figure}
}

\vspace{-0.6em}
\section{Additional Related Work}\label{sec:app:additional:relatedwork}
\vspace{-0.4em}

\hspace*{0.8\parindent}
\textbf{Deep Clustering (DC).} 
PARTY~\cite{DBLP:conf/ijcai/PengXFYY16} is a two-stage DC method 
that uses autoencoders (AEs) with sparsity prior.
DEC~\cite{DBLP:conf/icml/XieGF16} is a single-stage AE-based method that jointly learns latent embeddings and cluster assignments 
by minimizing the KL divergence between the model's soft assignment and an auxiliary target distribution.
IDEC~\cite{DBLP:conf/ijcai/GuoGLY17} further improves DEC by integrating DEC's clustering loss and AE's reconstruction loss.
DCN~\cite{DBLP:conf/icml/YangFSH17} adopts the K-means objective to help AEs learn K-means-friendly representations.
In~\cite{DBLP:conf/ijcai/JiangZTTZ17}, variational AEs are used to model the data generative procedure for DC.
Recently, adversarial fairness has also been incorporated for deep fair clustering~\cite{DBLP:conf/cvpr/LiZL20}.